\documentclass[useAMS,usenatbib]{mn2e}
\usepackage{graphicx,epsfig,lscape}
\usepackage{times,deluxetable}
\title[Spectral index properties of mJy radio sources]{Spectral index properties of milliJansky radio sources}
\author[Randall et al.]{K.~E. Randall$^{1,2}$\thanks{E-mail:
krandall@physics.usyd.edu.au}, A.~M. Hopkins$^{3}$, R.~P. Norris$^{2}$, P.-C. Zinn$^{4}$, E. Middelberg$^{4}$,\newauthor M.~Y. Mao$^{2,3,5}$, R.~G. Sharp$^{6}$\\
$^{1}$Sydney Institute for Astronomy, School of Physics, The University of Sydney, NSW 2006, Australia\\
$^{2}$CSIRO Astronomy and Space Science (CASS), P.O. Box 76, Epping NSW, 1710, Australia\\
$^{3}$Australian Astronomical Observatory, PO Box 296, Epping, NSW, 1710, Australia\\
$^{4}$Astronomisches Institut der Ruhr-Universit$\ddot{a}$t Bochum, Universit$\ddot{a}$tsstr. 150, 44801 Bochum, Germany\\
$^{5}$School of Mathematics and Physics, University of Tasmania, Private Bag 27, Hobart, Tasmania 7001, Australia\\
$^{6}$Research School of Astronomy and Astrophysics, The Australian National University, Cotter Road, Weston, ACT 2611, Australia}

\begin{document}

\date{Accepted 2011 December 20. Received 2011 December 16; in original form 2011 October 17}

\pagerange{\pageref{firstpage}--\pageref{lastpage}} \pubyear{2011}

\maketitle

\label{firstpage}
\begin{abstract}
At the faintest radio flux densities ($S_{1.4}<10$\,mJy), conflicting results have arisen regarding whether there is a flattening of the average spectral index between a low radio frequency (325 or 610\,MHz), and e.g. 1.4\,GHz. We present a new catalogue of 843\,MHz radio sources in the ELAIS-S1 field that contains the sources, their ATLAS counterparts, and the spectral index distributions of the sources as a function of flux density. We do not find any statistically significant evidence for a trend towards flatter spectral indices with decreasing flux density. We then investigate the spectral index distribution with redshift for those sources with reliable redshifts and explore the infrared properties. An initial sample of faint Compact Steep Spectrum sources in ATLAS is also presented, with a brief overview of their properties. 
\end{abstract}


\section{Introduction}
The desire to understand the properties of the faintest radio source populations has led to numerous surveys pushing to ever fainter levels. There are now a large number of deep and wide area radio surveys available, such as the Australia Telescope Large Area Survey \citep[ATLAS;][]{ray,Middelberg08}, the Cosmic Evolution Survey \citep[COSMOS;][]{scoville,smolcic}, the ATESP Survey \citep{prandoni,prandoni1,prandoni2}, the Phoenix Deep Survey \citep{phoenix}, and many others \citep{seymour,owen,owen1,ibar,ibar1}. The bright radio source population ($S_{1.4}>10$mJy) is well studied, and is predominantly composed of active galactic nuclei (AGN) \citep{condon,grup,maglio,georg,afonso1}. At fainter flux densities,  star-forming galaxies (SFGs) begin to dominate the radio source population, particularly into the $\mu$Jy regime \citep{windhorst85,srccnts,afonso,seymour}. Understanding these faint radio source populations is essential for understanding galaxy evolution, the role of star formation, AGN and the relationship between the two. The use of multiwavelength data to complement radio surveys is vital in order to distinguish between AGN and star formation processes as the origin of the radio emission down to the faintest flux densities, although there is likely also a composite population at these faint levels \citep{hill99,hill01}. The relative proportions of these two populations will have an effect on the average radio spectral index ($\alpha$)\footnote{Assuming $S_{\nu}\propto\nu^{\alpha}$, where S is the measured flux density and $\nu$ is the observer's frame frequency.} as a function of flux density, and at the faintest flux densities,  core-dominated AGN may be more prevalent, and may flatten the average spectral index to $\alpha>-0.7$. 

Conflicting evidence has arisen over the nature and properties of sources at frequencies below 1.4\,GHz, particularly of their spectral index properties, and whether there is a flattening of the average spectral indices for faint ($S_{1.4}<10$mJy) radio sources. \citet{prandoni2,prandoni4} found sources with fluxes less than a few milliJansky had an average spectral index which was flatter than that of the brighter radio sources ($\alpha\sim-0.7$). Similarly, \citet{owen} found a flattening of the average spectral index between 325\,MHz and 1.4\,GHz for 1.4\,GHz selected radio sources with $S_{1.4}<10$mJy, and angular sizes $>3''$. They found though, that the spectral indices steepen again at the faintest flux density end of the 20\,cm survey ($\sim\,0.5\,$mJy). In contrast, in the deepest radio field, the Lockman Hole \citep{ibar}, no flattening of the spectral indices between 610\,MHz and 1.4\,GHz was seen for flux densities S$_{1.4}>100\mu$Jy.

A flattening of the average spectral indices implies that there is a flat or inverted spectrum population of sources at these milliJansky flux densities, at fluxes fainter than where the steep spectrum star-forming population emerges ($\sim0.5\,$mJy). Investigating the milliJansky and microJansky radio source populations at different frequencies allows us to investigate this suspected flattening of the average spectral indices, and the population of sources causing the flattening, particularly if we can fill in the frequency regime from very low frequencies ($\sim100$\,MHz) up to 1.4\,GHz. It is necessary to understand the spectral index properties of these faint radio sources for many reasons, such as the  $z$-$\alpha$ relation, used to find the most distant radio galaxies \citep{debreuck04,klamer,ishwara}, identifying young radio AGN, such as Gigahertz Peaked Spectrum (GPS) and Compact Steep Spectrum (CSS) sources \citep{Review,young,morganti} or determining the emission mechanism in radio galaxies, whether it is from star formation, or the central AGN \citep{prandoni2}.

Here we present radio observations at 843\,MHz in the ELAIS-S1 field, for which we have complementary 1.4 and 2.3\,GHz data from ATLAS that we can use to investigate the possible spectral index flattening at a frequency nearer to the ubiquitous 1.4\,GHz radio surveys. Section~\ref{sec:elais} briefly reviews the existing multiwavelength data covering the ELAIS-S1 region and ATLAS, and in Section~\ref{sec:data} we describe our observations, the data reduction, and the cross-matching process to the ATLAS 1.4 and 2.3\,GHz catalogue. We present the catalogue in Section~\ref{sec:catalogue}. Our results and analysis are explored in Section~\ref{sec:results}, and a new faint sample of candidate CSS sources is presented in Section~\ref{sec:props}, discussing the initial selection and properties of these sources. Our results and plans for future work are discussed in Section~\ref{sec:discussion} and our conclusions are presented in Section~\ref{sec:concl}. Throughout this analysis, unless otherwise noted, we use the cosmological parameters, $\Omega_{M}=0.27$, $\Omega_{\Lambda}=0.73$ and $H_0=71$\,kms$^{-1}$Mpc$^{-1}$ \citep{wmap}.

\section{The ELAIS-S1 Region}
\label{sec:elais}
The European Large Area ISO Survey - South 1 Region (ELAIS-S1) has been a target for many multiwavelength surveys and targeted observations over the last decade. The ELAIS-S1 field covers $2^{\circ}\times2^{\circ}$, centred on RA = $00{^h}34{^m}44^{s}.4$ and Dec=$-43^{\circ}28'12''.0$ (J2000.0). Observations covering the ELAIS-S1 field include radio imaging, optical imaging and spectroscopic redshifts, infra-red observations (near, mid and far-infrared), UV, and X-ray observations. Part of the attraction for the multiwavelength surveys in this region is that this field has the lowest Galactic 100$\mu$m cirrus emission in the southern sky \citep{schlegel}, including the absolute minimum. 

\subsection{Multiwavelength data in ELAIS}
\subsubsection{ISO}
The field was first observed as part of a deep, wide-angle survey with the Infrared Space Observatory (ISO) at 15 and 90\,$\mu$m \citep{oliver}. Follow-up observations were done at 6.7\,$\mu$m, and the catalogue covering the S1 Region at these three wavelengths is discussed in \citet{rowan}. The complete 15\,$\mu$m catalogue contains all the sources from the ELAIS regions: N1, N2, N3, S1, S2, and was finalized by \citet{vaccari}. The three bands of this survey have rms noise levels of 1.0, 0.7 and 70 mJy respectively for the 6.7, 15 and 90\,$\mu$m bands. Photometric uncertainties for all bands were $\sim10\%$, and the astrometric accuracy is $\sim0.5''$.
\subsubsection{Spitzer/SWIRE}
\label{sec:swire}
The ELAIS-S1 field was observed by the Spitzer Space Telescope as part of the largest Legacy Program, the Spitzer Wide-area Infra-Red Extragalactic survey \citep[SWIRE;][]{lonsdale,lonsdale1}. SWIRE aimed to trace the evolution of dusty SFGs, AGN, and evolved stellar populations out to a redshift of $z\sim$3, by imaging large areas of sky in seven different infrared bands. In combination with optical imaging, this would allow SED modeling and exploration of galaxy evolution with environment.
The seven imaging bands of SWIRE used two different instruments aboard Spitzer, the Multiband Imaging Photometer \citep[MIPS;][]{rieke}, and the Infrared Array Camera \citep[IRAC;][]{fazio}, covering $\sim7$\,deg$^2$ over the ELAIS-S1 region. The 24\,$\mu$m MIPS imaging reaches a 5$\sigma$ sensitivity of 350\,$\mu$Jy, and the IRAC bands of 3.6, 4.5, 5.8 and 8.0\,$\mu$m have 5$\sigma$ sensitivities of 3.7, 5.3, 48 and 37.75\,$\mu$Jy respectively. 
\subsubsection{ESIS}
The ESO-Spitzer Imaging extragalactic Survey (ESIS) is the optical follow-up to SWIRE, consisting of optical imaging in the B, V, R bands with the Wide Field Imager \citep[WFI;][]{berta}, and I and z band imaging with VIMOS \citep{berta1}. Currently, the VIMOS observations have been completed, but the WFI observations are not yet fully processed. The BVR observations described in \citet{berta} cover 1.5\,deg$^2$, in the central region of the ELAIS-S1 field. The catalogue of BVR sources is 95\% complete to 25$^{m}$ in B and V, and 24.5$^{m}$ in R. 132712 sources are included in this catalogue, with an rms uncertainty of $\sim0.15''$ for the coordinates of the sources. The VIMOS data covers $\sim4$\,deg$^2$ in the I band and $\sim1$\,deg$^2$ in the z band, resulting in a completeness of 90\% at 23.1$^{m}$ in the I band, and 22.5$^{m}$ in the z band. Over 300,000 sources were catalogued in the I band, and over 50,000 in the z band, with an rms of $\sim0.2''$ in both bands.
\begin{figure*}
\begin{center}
\includegraphics[scale=0.6]{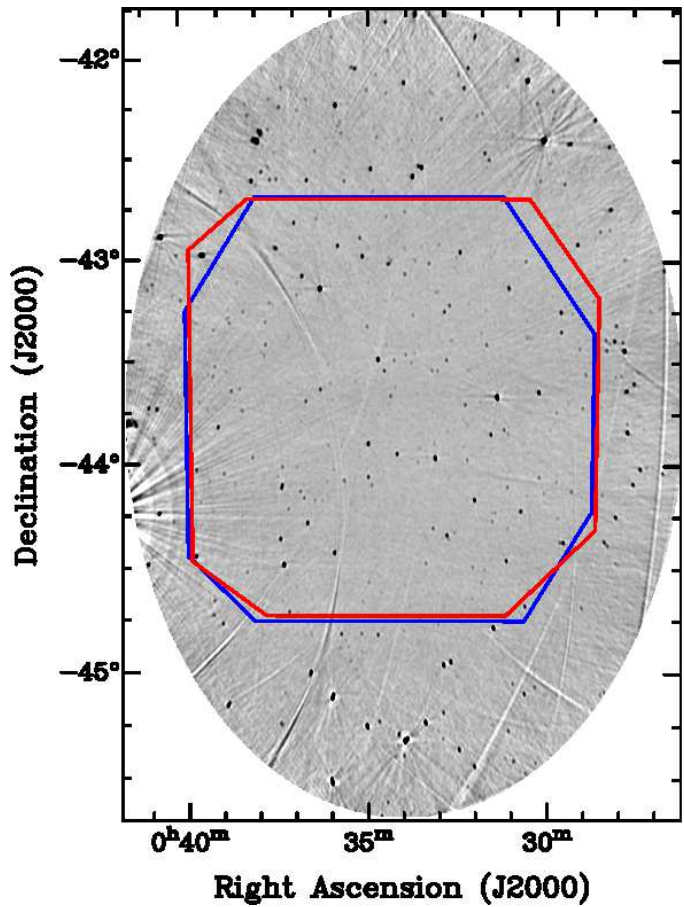}
\caption{The greyscale MOST 843\,MHz image, with overlays indicating the approximate borders of the ATCA 1.4\,GHz and 2.3\,GHz observations. The larger, red boundary represents the 1.4\,GHz mosaic. 
\label{fig:mostimage}}
\end{center}
\end{figure*}

\subsubsection{Previous Radio Observations}
Prior to the ATLAS observations (detailed below in \S\ref{sec:atlas}), \citet[][G99]{grup} observed the ELAIS-S1 field using the Australia Telescope Compact Array (ATCA) in 1997, covering $\sim7$\,deg$^2$. The observations consisted of a mosaic of 49 different pointings, resulting in an image rms of $\sim80\mu$Jy at 1.4\,GHz,  and a catalogue of 581 radio sources. The lowest flux density sources were catalogued down to a 5$\sigma$ level of 0.2\,mJy in the centre of the field, and 0.4\,mJy in the remaining area.

\subsubsection{X-ray Observations}
BeppoSAX, an X-ray satellite, first observed 40\% of the ELAIS-S1 field in 1999 \citep{alexander}. These observations reached a sensitivity of $\sim10^{-13}$\,erg\,cm$^{-2}$s$^{-1}$ in the 2$-$10\,keV range, covering $\sim1.7$\,deg$^2$. The central $\sim0.6$\,deg$^2$ were then surveyed by XMM-Newton in four deep pointings \citep{puccetti}. The XMM observations were taken in both soft and hard X-rays, and each pointing had a net exposure time of $\sim60$\,ks. A total of 478 sources were detected, with 395 in the soft X-ray band (0.5$-$2\,keV) and 205 in hard X-ray (2-10\,keV). The flux limits are  $\sim5.5\times10^{-16}$\,erg\,cm$^{-2}$s$^{-1}$ and $\sim2\times10^{-15}$\,erg\,cm$^{-2}$s$^{-1}$ respectively for the soft and hard bands. 

\subsubsection{UV Observations}
The ultraviolet Galaxy Evolution Explorer \citep[GALEX;][]{martin} observed the ELAIS field in two of its observing modes; the Deep Imaging and Wide Spectroscopic Survey modes \citep{burgarella}. The bands observed were centred on 153 and 231\,nm, and the current data release (GR6\footnote{http://galex.stsci.edu/GR6/}) covers 70\% of the sky in the imaging mode, and has 61439 spectroscopic sources. The ELAIS field was observed for $\sim9$\,hours in the spectroscopic survey mode, and $\sim3$\,hours in the imaging mode. 

In the analysis below we focus on the infrared data as the primary complement to our radio data, although in future work we will also take advantage of the X-ray and UV measurements.

\subsection{The Australia Telescope Large Area Survey (ATLAS)}
\label{sec:atlas} 
ATLAS is the widest, deep radio survey to date \citep[][N06, M08]{ray,Middelberg08}, covering $\sim7$deg$^2$ over two fields, ELAIS-S1, and the Chandra Deep Field South (CDFS). The rms of the 1.4\,GHz imaging data is currently 30\,$\mu$Jy, and the aim is to achieve an rms of 10\,$\mu$Jy across both fields \citep{banfield}. ATLAS observations have been completed at 1.4 and 2.3\, GHz with ATCA from 2006 to 2010, with one data release in 2008, and subsequent data releases to begin in late 2011 \citep{chales,banfield}. ATLAS has many scientific goals, primarily focussed on investigating the evolution of galaxies and AGN. Specific goals include distinguishing between AGN and SFGs and determining the contribution of each to a given galaxy's luminosity, searching for high-$z$ radio galaxies to trace the formation of clusters at high redshift \citep{minnie}, and finding new types of rare sources \citep[N06, M08,][]{ifrs}. The ATLAS survey regions were chosen because of the large amount of multiwavelength data covering these two fields, including near and far-infrared, and deep optical data, and in some areas, X-ray and UV. We aim to create the most comprehensive multiwavelength survey of faint radio sources to date. ATLAS currently contains $\sim\,2000$ radio sources, and we estimate we will have $\sim\,16000$ radio sources following final analysis of new observations completed in 2010 from ATCA with the Compact Array Broadband Backend \citep[CABB;][]{cabb}. These new observations consist of 1000 hours of integration time, and cover a bandwidth of 0.5\,GHz centred on 1.4\,GHz. 

In ELAIS-S1, 1276 radio sources have been catalogued, comprising 1366 radio components, at 1.4\,GHz \citep[][M08]{Middelberg08}. The 2.3\,GHz image has a lower sensitivity, with an rms noise level of $\sim60\mu$Jy in the central $\sim1$deg$^2$ and $\sim100\mu$Jy over the entire field. We find only 576 radio sources have a well-defined counterpart at 2.3\,GHz, mainly because the resolution is $\sim3$ times coarser at this frequency than at 1.4\,GHz as a consequence of the compact configuration of ATCA used for these observations. For the purposes of this paper and catalogue, we use the cross-matches by \citet[][Z11]{zinn}, where we take the 1.4\,GHz sources matched to the poorer resolution 2.3\,GHz data, and extract the fluxes from these images for each source. M08 also compared their flux densities and positions to G99 by repeating their source extraction on the G99 1.4\,GHz radio image. The differences in radio positions were determined to be negligible, but the flux densities of M08, while within $\sim\,3$\% of the earlier measurements, were consistently higher than G99. 
\begin{figure*}
\begin{center}
\includegraphics[scale=0.28]{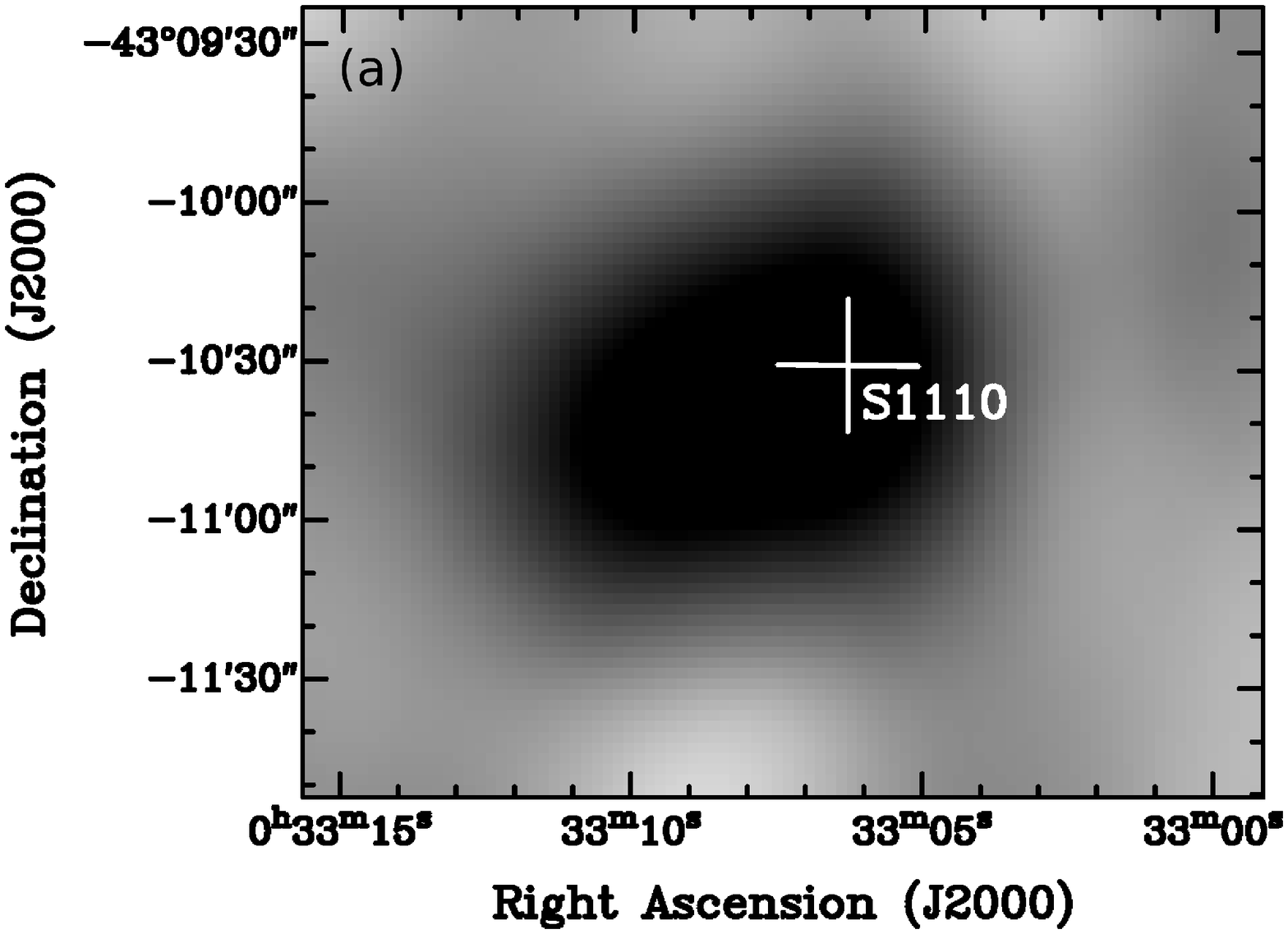}\hfill \includegraphics[scale=0.28]{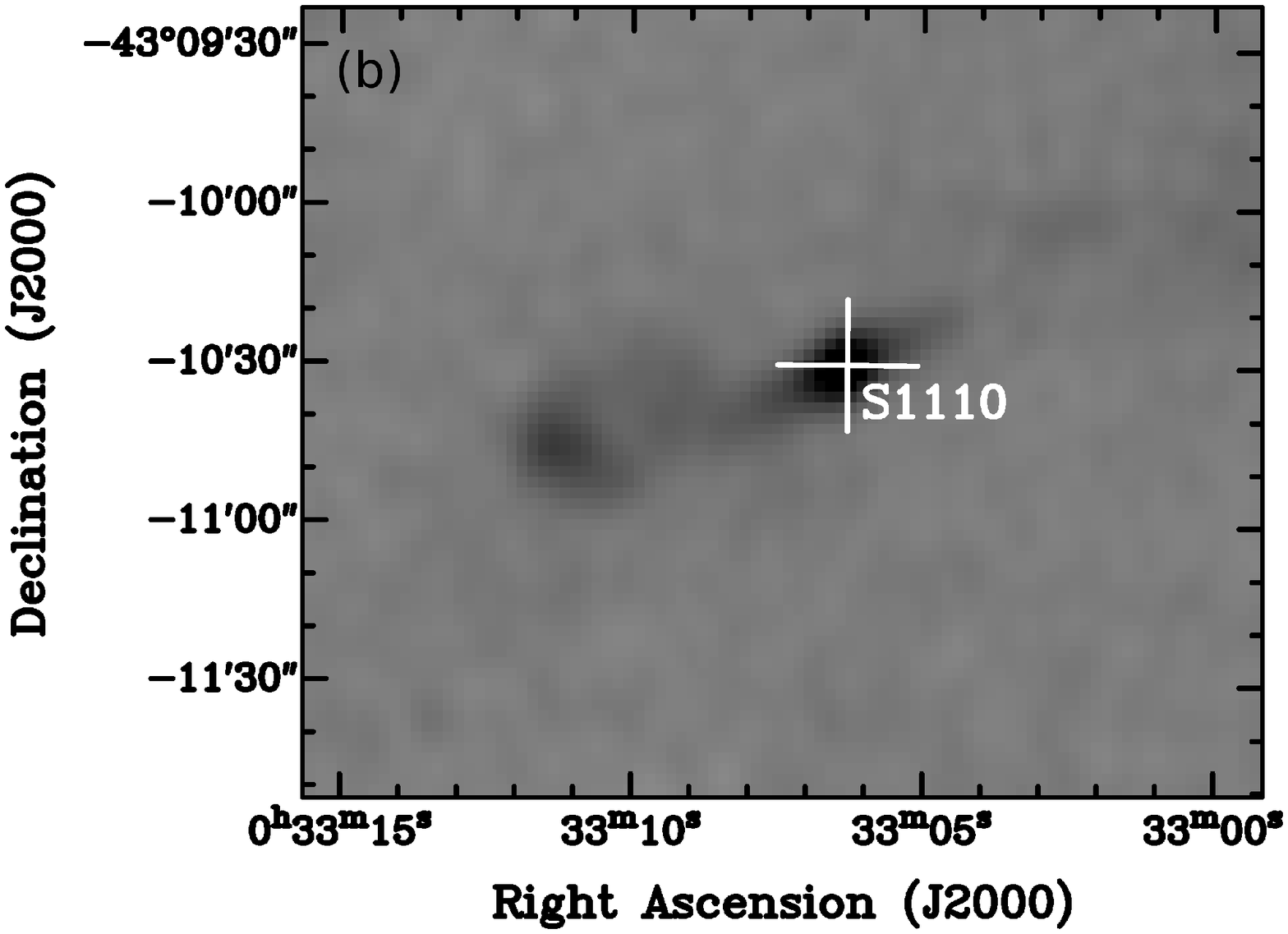}\hfill \includegraphics[scale=0.28]{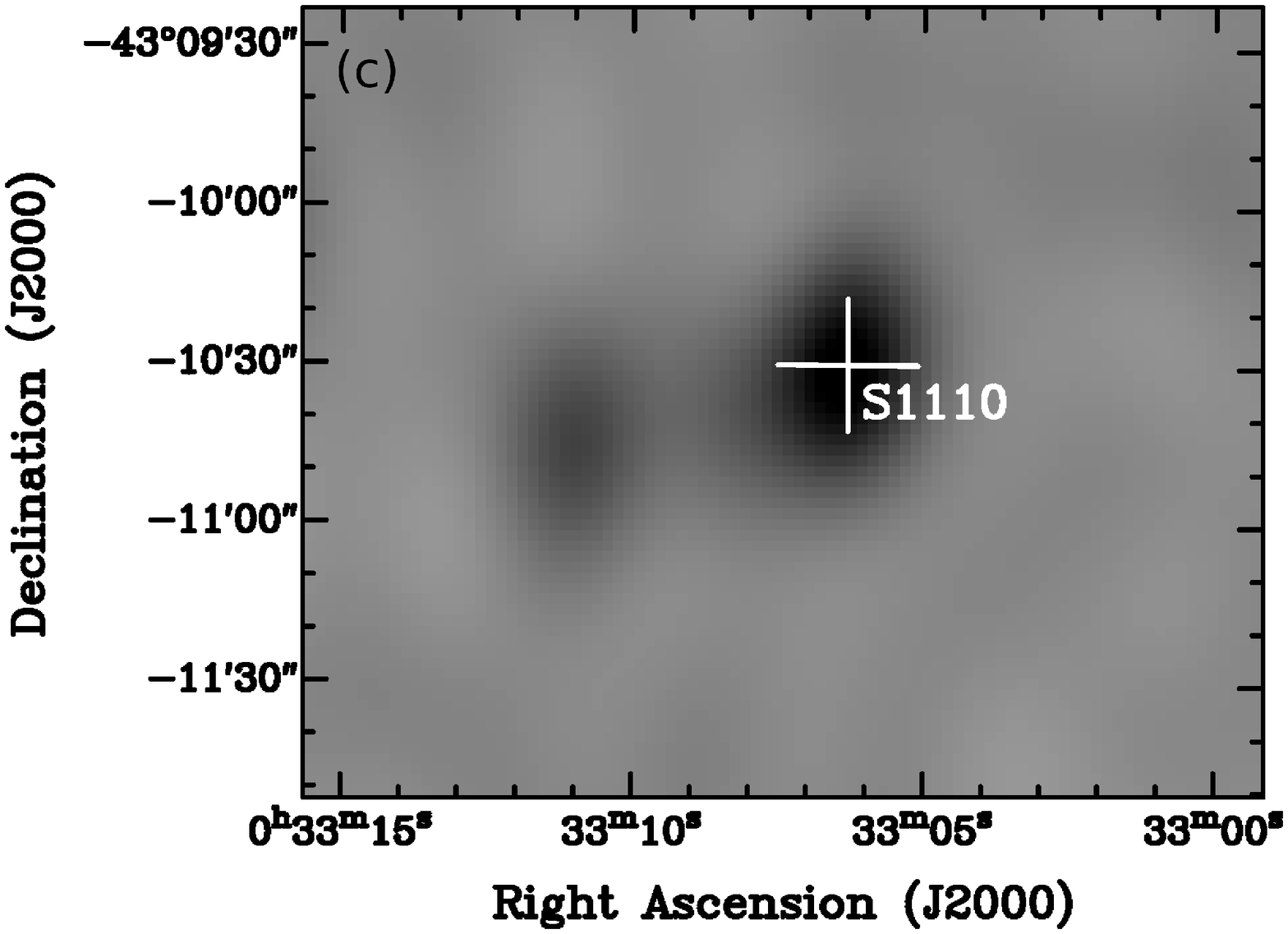}
\caption{(a) Greyscale MOST 843\,MHz image of ATELAIS J003306.30-431029.8, (b) original ATCA 1.4\,GHz greyscale image, and (c) the original ATCA 2.3\,GHz greyscale image. The white cross indicates the 1.4\,GHz radio position of this source. 
\label{fig:resol}}
\end{center}
\end{figure*}

\begin{figure*}
\begin{center}
\includegraphics[scale=0.28]{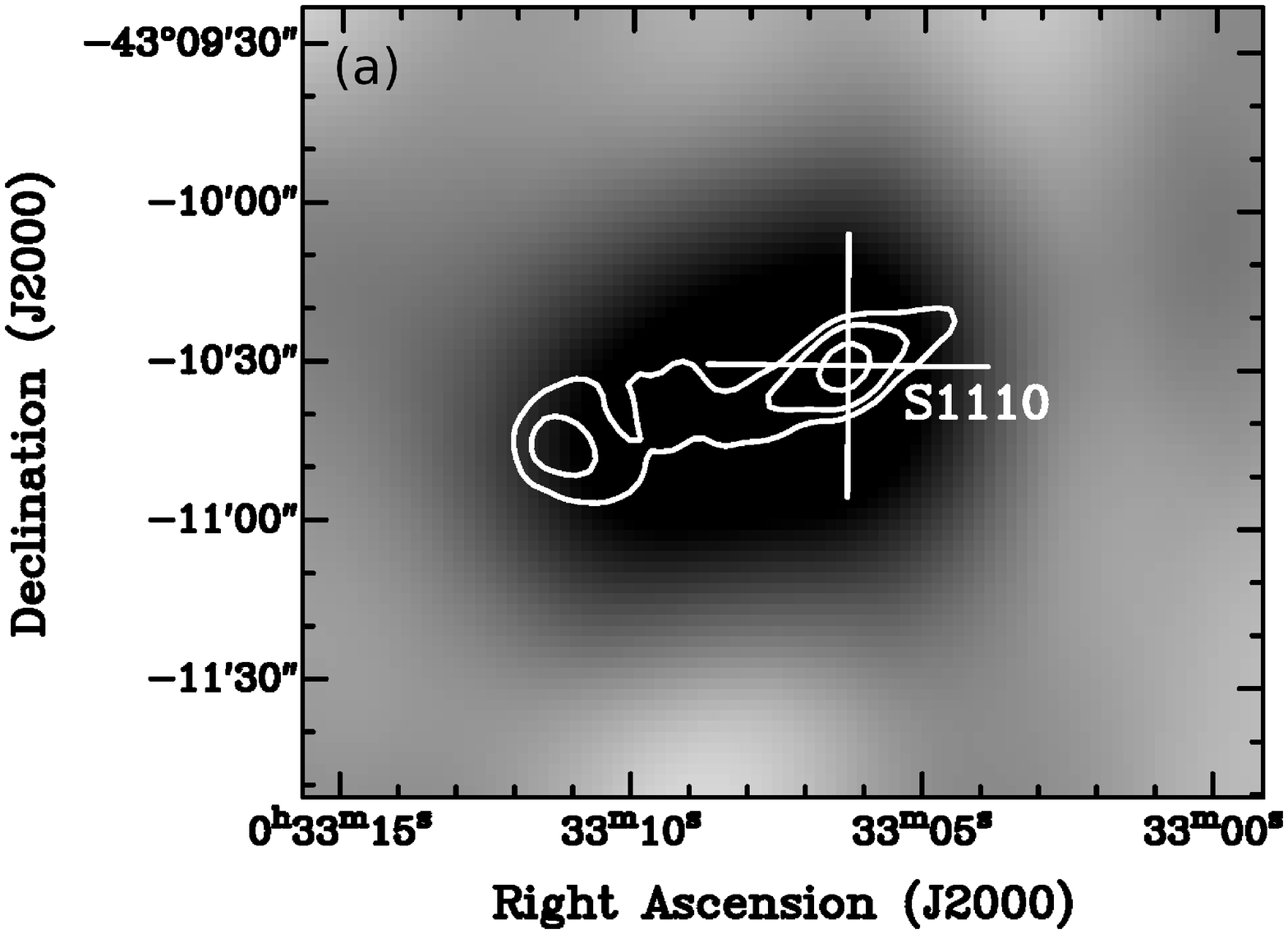}\hfill \includegraphics[scale=0.28]{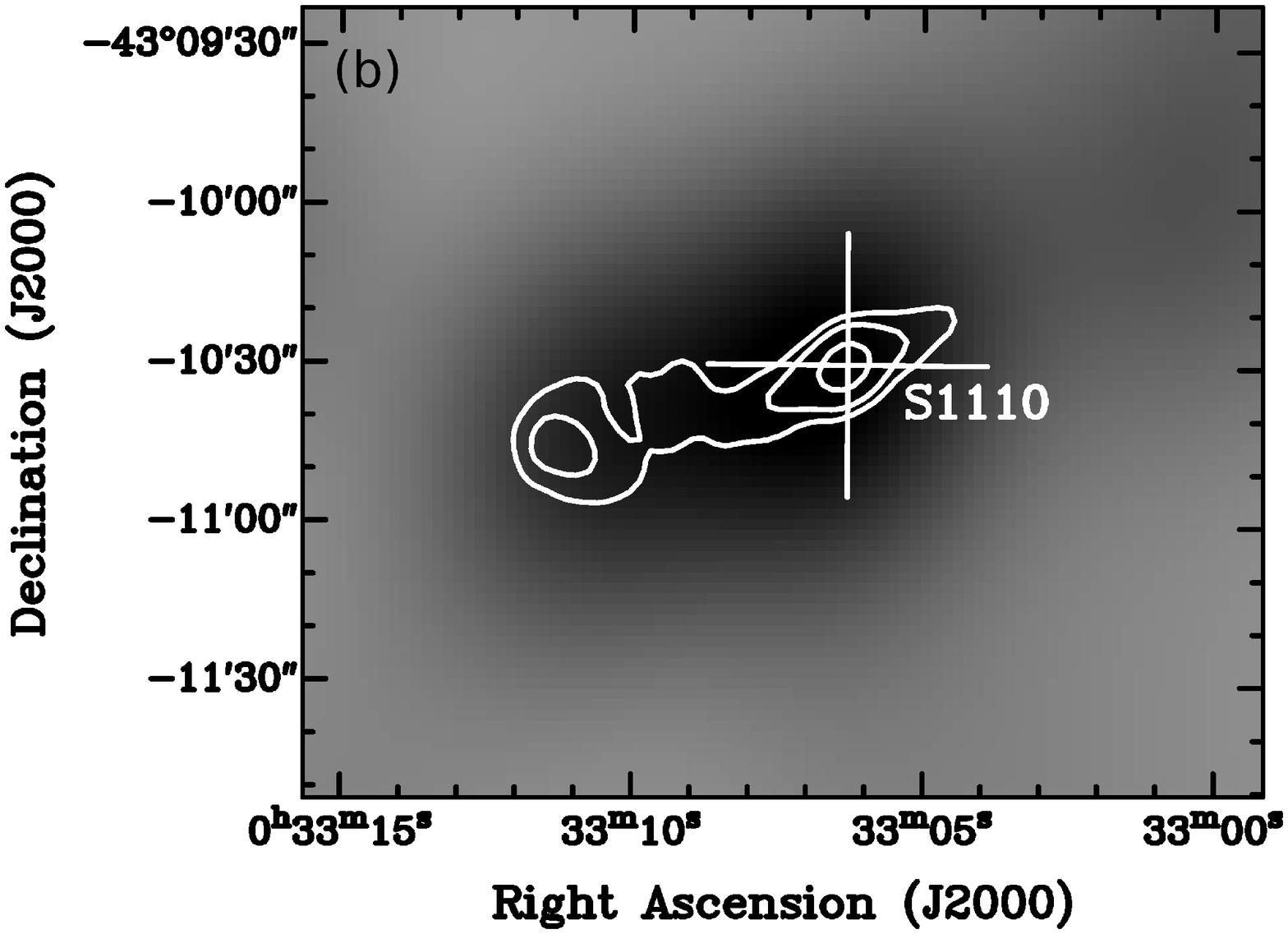}\hfill \includegraphics[scale=0.28]{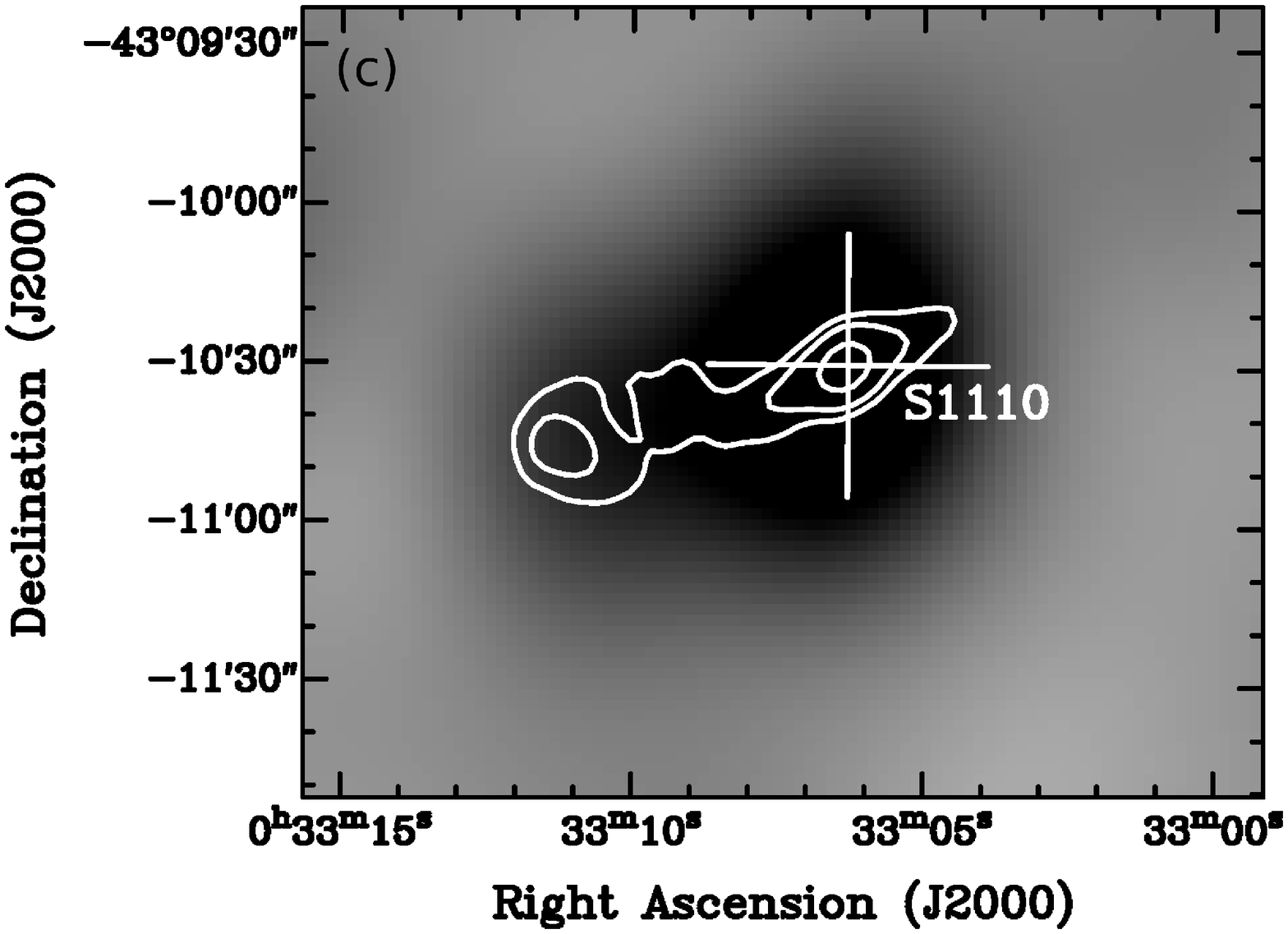}
\caption{(a) The greyscale MOST 843\,MHz image of ATELAIS J003306.30-431029.8, (b) the convolved ATCA 1.4\,GHz greyscale image, and (c) the convolved ATCA 2.3\,GHz greyscale image. The white cross indicates the 1.4\,GHz radio position of this source, and the 1.4\,GHz contours are overlaid at levels of 2.5, 5 and 12\,mJy
\label{fig:resol1}}
\end{center}
\end{figure*}

\section{Observations, data reduction and cross-matching}
\label{sec:data}
To obtain low-radio frequency (843\,MHz) data within ATLAS, we used the Molonglo Observatory Synthesis Telescope \citep[MOST; ][]{mills,robertson}. The MOST is a 1.6\,km long cylindrical paraboloid reflector, with its axis aligned in an east-west direction. It operates at a frequency of 843\,MHz, and has a field of view of $2{\fdg}7cosec(\delta)\times2{\fdg}7$deg$^2$ \citep{large,bock}. We have 31 separate 12 hour observations taken with MOST, which were combined into a single image. The CDFS field (centred at RA = $03{^h}32{^m}28^{s}$ and Dec=$-27^{\circ}48'31\farcs0$ (J2000.0)) was not observed, due to strong radio frequency interference (RFI) that increased with lower elevation.
\subsection{Processing}
The data reduction pipeline for MOST data is highly automated and reliable. Unfortunately, the telescope was subject to severe RFI at the time of our observations. The process to remove the RFI required manual identification and excision of the affected data. Once completed, we were able to re-run the data pipeline, and produce final cleaned images. Seven of our 31 observations were subject to RFI that we were unable to remove completely, and these were not included in our final imaging. The remaining 24 images were added together in MIRIAD, first by re-gridding the images to ensure a common astrometry and pixel grid, and then using the task \textsc{imcomb} to combine the images.  The final image has an rms sensitivity of $\approx0.6$\,mJy, a higher value than expected due to the presence of low-level RFI that could not be excised. We catalogue sources down to a $5\sigma$ level of 3\,mJy. The final image is shown in Figure~\ref{fig:mostimage} with the approximate borders of the ATLAS 1.4 and 2.3\,GHz mosaics overlaid.

The MOST image was subject to two kinds of artifacts, grating rings and radial spokes. Due to the nature of the telescope design and observing mode, these artifacts are technically difficult to remove. The artifacts remain in our image, and due care was taken to ensure all objects in our final catalogue were not spurious sources associated with the artifacts. An in-depth discussion of these artifacts is given in \citet{SUMSS2}.
\subsection{Source-finding}
Source finding for the final MOST image was done using the task \textsc{sfind} \citep{sfind} in MIRIAD. This produces a catalogue of source positions, peak and total flux density and errors, plus the attributes of the Gaussian fits for each source. The final catalogue contains 325 radio sources, after removal of $\approx100$ spurious sources associated with the grating rings and radial spokes. The spurious sources were identified and removed by visual inspection (most lay directly on a strong radial spoke or diffraction ring), and by comparison to the Sydney University Molonglo Sky Survey \citep[SUMSS;][]{bock,SUMSS2} images (see \S\ref{sec:fluxdist} for details). Each MOST ATLAS source was identified in the SUMSS images with a low signal-to-noise source that was below the detection level of the SUMSS catalogue. The MOST image is larger than the ATLAS survey region, so our catalogue encompasses a larger number of sources in total than we subsequently analyse together with the other ATLAS data. 

We use the process outlined in \citet{phoenix} to calculate the errors associated with our measured flux densities, and use Equation 5 of \citet{phoenix} with one small modification:\\
\begin{equation}
\sigma_I=I \sqrt{2.5 \frac{\sigma^2}{I^2} + 0.05^2},
\label{eq:error}
\end{equation}
where $I$ is the total integrated flux density, $\sigma$ is the rms error in the image at the source location, and $\sigma_I$ is the total error on the integrated flux density. We use $0.05^{2}$ as the sum of the squares of the instrumental and pointing errors as given in M08, instead of $0.01^{2}$ as given in \citet{phoenix}. We have used the more conservative estimate of these errors as given by M08, due to the calibration accuracy of MOST being comparable to that of ATCA.

\subsection{Cross-matching to ATLAS}
\label{sec:crossmatch} 
The final MOST catalogue was positionally cross-matched to the combined 1.4 and 2.3\,GHz catalogue produced by Z11, that included the relevant SWIRE and optical data. As mentioned previously in \S\ref{sec:atlas}, only 576 of the 1276 1.4\,GHz radio sources in ELAIS have a reliable single 2.3\,GHz counterpart, whilst 460 have no counterpart, and 240 are blended or poorly fitted sources (where the 2.3\,GHz counterpart encompasses multiple 1.4\,GHz sources). From our cross-matching, 105 MOST sources were matched to single sources in the Z11 catalogue. Another 30 MOST sources were matched to confused or blended sources from Z11, and 31 MOST sources were found to have multiple isolated Z11 sources matched to a single MOST source. 
\subsection{Resolution Matching}
\label{sec:convolution} 
The resolution of the MOST image, $62''\times\,43''$, is much coarser than the resolution of the 1.4\,GHz image ($10''\times7''$), and the 2.3\,GHz image ($33''\times20''$). To determine accurate spectral indices across the three frequencies, it is necessary to convolve the ATLAS 1.4 and 2.3\,GHz images to the same size as the MOST beam to ensure the recovered flux for sources at all frequencies includes any emission extended on scales up to those consistent with the MOST beam. Figure~\ref{fig:resol} shows an example of a complex source from the ATLAS 1.4\,GHz catalogue in greyscale in the three radio frequencies, to highlight the differences in resolution. In Figure~\ref{fig:resol1}, the same source is shown after convolving the 1.4 and 2.3\,GHz images to the same resolution as the MOST image. Although convolution removes most of the small scale structure visible in the original ATLAS 1.4\,GHz image, it ensures we are detecting emission from the same spatial region from each source.

For consistency, we used \textsc{sfind} to detect sources within the convolved 1.4 and 2.3\,GHz images. This allows us to measure errors on our flux densities that are consistent for the three images, which is important  for producing accurate spectral indices and associated uncertainties. Rather than rejecting blended sources entirely, we account for them carefully in our catalogue. These are single or point sources in the 843\,MHz MOST image (and the 1.4 and 2.3\,GHz convolved ATLAS images) that encompass multiple ATLAS 1.4\,GHz sources from the M08 catalogue (Figure~\ref{fig:complex}). The M08 flux densities for these sources reported in our catalogue are the summed values of the individual M08 1.4\,GHz flux densities. As the contribution to the total flux density from each individual component within a blended source cannot be determined,  we treat them as a single entity for the purpose of the current analysis, even though they may be physically unrelated. We include these sources in our analysis for completeness, but caution that their spectral indices should not be considered as accurate estimates of those for the underlying source components. 
\subsubsection{Flux Density Comparisons and Corrections}
To ensure our measured 1.4 and 2.3\,GHz flux densities from the convolved, coarser-resolution, images were robust, we compared these to the flux densities in M08 and Z11. We find that our measurements in the images convolved to a resolution consistent with the MOST image tend to systematically overestimate the 1.4\,GHz flux densities compared to M08, and underestimate the 2.3\,GHz flux densities compared to Z11.

The issue at 2.3\,GHz arises from negative CLEAN bowls around sources in the original image. These artifacts give rise, after convolution, to our observed systematic decrease in flux density. The 2.3\,GHz beam size ($33''\times20''$) is close to that of our MOST images ($62''\times\,43''$), and the bulk of the sources are unresolved, meaning that there is likely to be little flux missed on extended scales. Consequently, we choose to use the flux densities estimated by Z11 in the original 2.3\,GHz image for our spectral index estimates.

At 1.4\,GHz, the difference in resolution from the MOST image is sufficient that we need to use the flux densities from the convolved image. The flux density overestimate compared to M08, which is limited to the fainter sources ($S_{843}<10$\,mJy), is typically of the order of 10\%. This exists for clearly unresolved sources, which should be identical before and after convolution. This issue is associated with side-lobes from the radio sources in each individual telescope pointing, below the level to which the image has been CLEANed, being summed in the convolved image. While this is a small effect, with a minimal impact on our derived spectral indices (quantified below), we can make an empirical correction that minimizes any impact further still. This is implemented through a least-squares fit of our convolved flux densities against those of M08, and scaling our convolved flux densities using this fit to be consistent with those of M08. This accounts for the flux-density dependence of the imaging systematics while retaining contributions from any real extended flux components, and at the same time, minimizing any possible overestimate of the flux density. This correction typically results in a change of only $\alpha_{fit}\sim0.02$.

The impact on our derived two-point spectral index estimates of potential remaining flux density uncertainties at 1.4 and 2.3\,GHz of $\sim10\%$ is $\sim$0.1. This uncertainty is included in our estimates of spectral index errors below, by being added in quadrature with the flux density errors (given by Equation~\ref{eq:error}).

\subsubsection{Blended sources}
Another effect of the convolution is that most objects classified as radio doubles, triples, or core-jet morphology by M08, appear as one MOST object (single or blended). Only two core-jet morphology sources appear slightly elongated in the direction of the jet in the MOST and convolved 1.4 and 2.3\,GHz images. 

Due to the lower resolution of the MOST image, there are 61 instances where there are ATLAS sources within the beam for a MOST radio source. These 61 MOST sources correspond to a total of 144 ATLAS M08 1.4\,GHz sources. Of the 61, 30 sources were classified as blended by Z11 due to having multiple 1.4\,GHz sources within the 2.3\,GHz beam ($33''\times20''$). The remaining 31 are classified as blended in this paper due to multiple 1.4 or 2.3\,GHz sources within the MOST beam. The cross-matched 1.4\,GHz ATLAS source (positionally closest to the MOST radio position) is generally bright while the second, blended, source contributing to the flux density is typically much fainter ($S_{1.4}<1$\,mJy). As a consequence, the majority of such blended sources are likely to have little contribution to the MOST flux density from the secondary component, and the spectral index estimate for the blended sources can be considered as that for the primary ATLAS counterpart, even though the uncertainties on this estimate will clearly be larger than the formal uncertainties provided in our catalogue. Blended sources are flagged in the catalogue. 

\begin{figure}
\begin{center}
\includegraphics[scale=0.35]{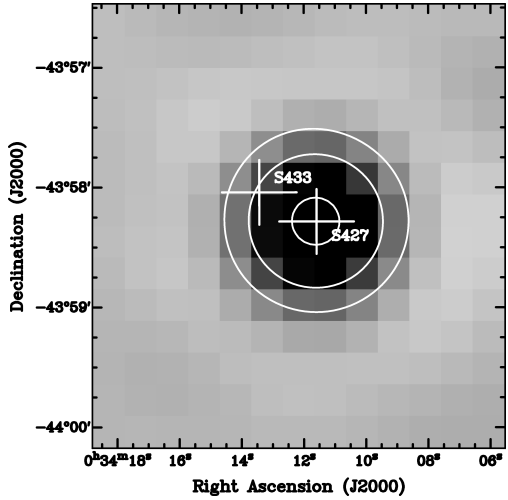}
\caption{An example of a blended source. The contours are from the convolved 1.4\,GHz image (at levels of 5, 10 and 20\,mJy), and the greyscale is the 843\,MHz image. The white crosses indicate the 1.4\,GHz radio position of the two sources, where the object in the centre of the image is the MOST cross-matched source, and the blended object indicated by the other cross. 
\label{fig:complex}}
\end{center}
\end{figure}

\begin{figure}
\begin{center}
\includegraphics[scale=0.35,angle=-90]{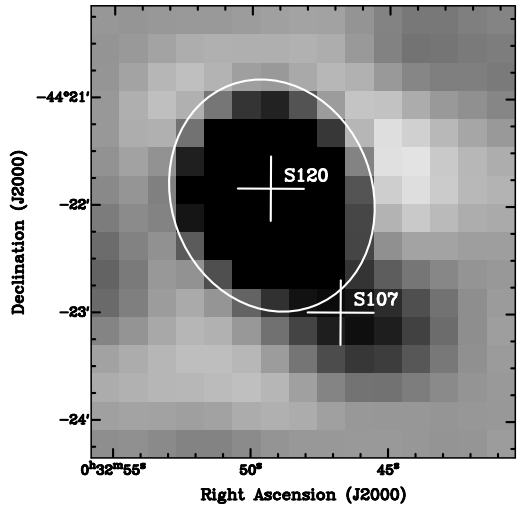}
\caption{The single example of a blended source where the ATLAS 1.4\,GHz radio position is not within the Gaussian (shown by the white ellipse) fitted to this source, but there is clearly contamination from the neighbouring source. 
\label{fig:complex1}}
\end{center}
\end{figure}

For blended sources in our convolved images, which were not classified as blended sources by Z11, the flux densities from separate sources in the ATLAS Z11 catalogue are summed. A source is defined as being part of a blended object if the 1.4\,GHz radio position lay within an ellipse the size of the Gaussian used to find the total flux density in the convolved image. An example of a blended source is shown in Figure~\ref{fig:complex} where the cross-matched ATLAS counterpart is in the centre of the MOST source, and the blended object is within the MOST synthesized beam. There is one exception to this, shown in Figure~\ref{fig:complex1}, where the blended object is outside the Gaussian fit, but the radio emission clearly extends from S120 into S107. 

\begin{figure*}
\begin{center}
\includegraphics[scale=0.4]{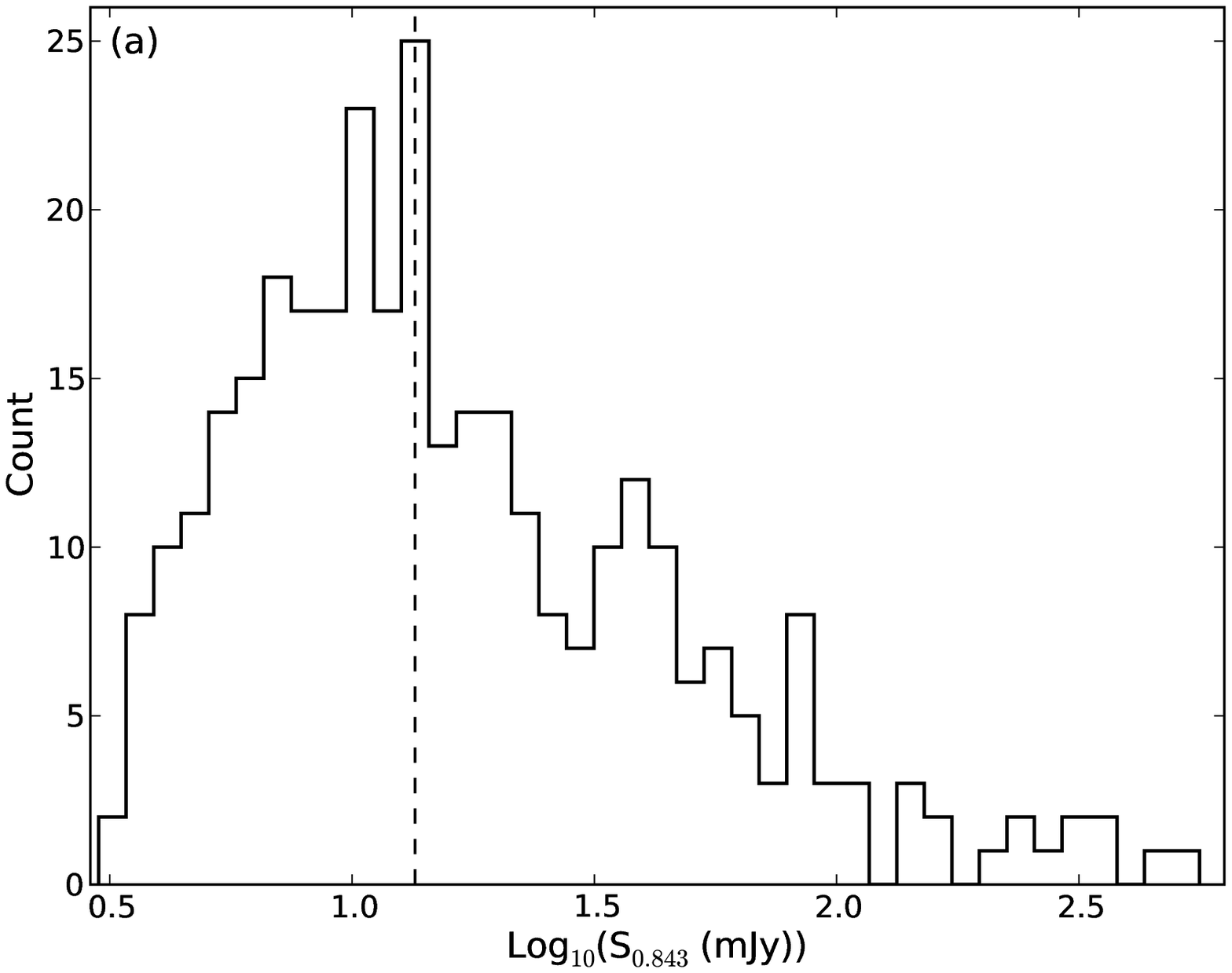}\includegraphics[scale=0.4]{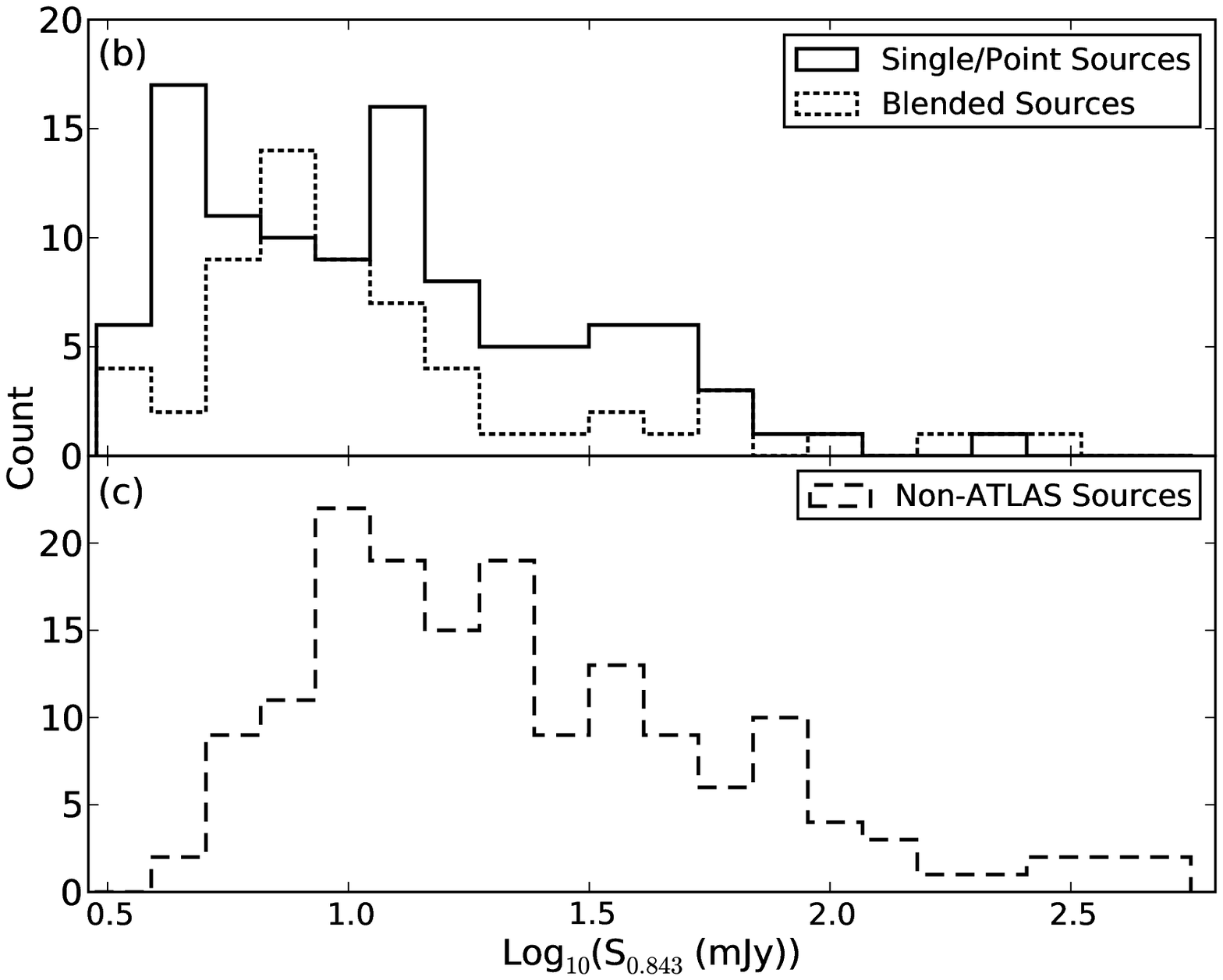}
\caption{(a) The flux density distribution for our entire catalogue of MOST sources with the median of the catalogue as the vertical dashed line. (b) Distribution of flux densities for single and blended MOST ATLAS sources, and (c) the MOST sources outside the observed ATLAS ELAIS field.
\label{fig:fluxes}}
\end{center}
\end{figure*}  

\begin{figure*}
\begin{center}
\includegraphics[scale=0.3]{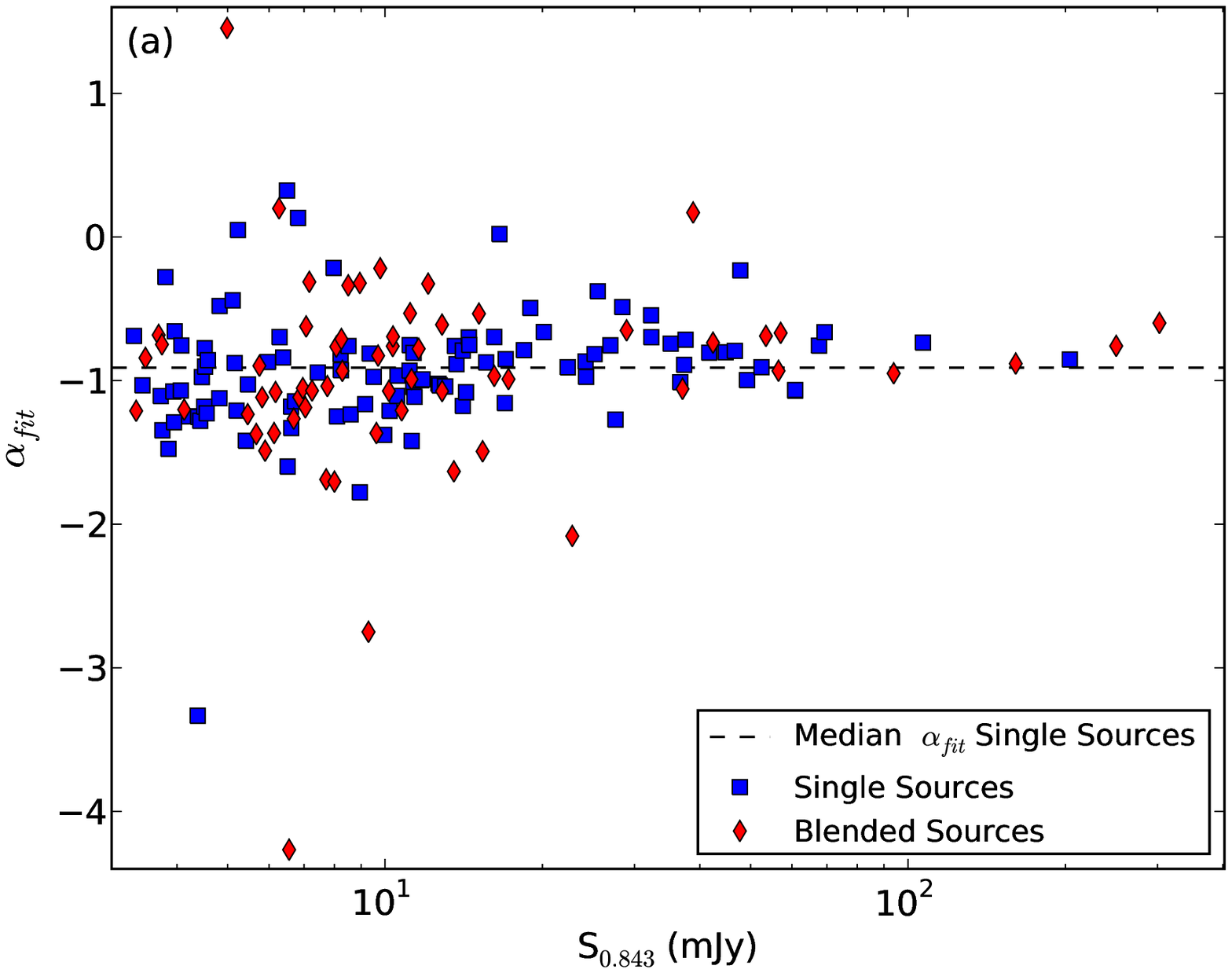}\includegraphics[scale=0.3]{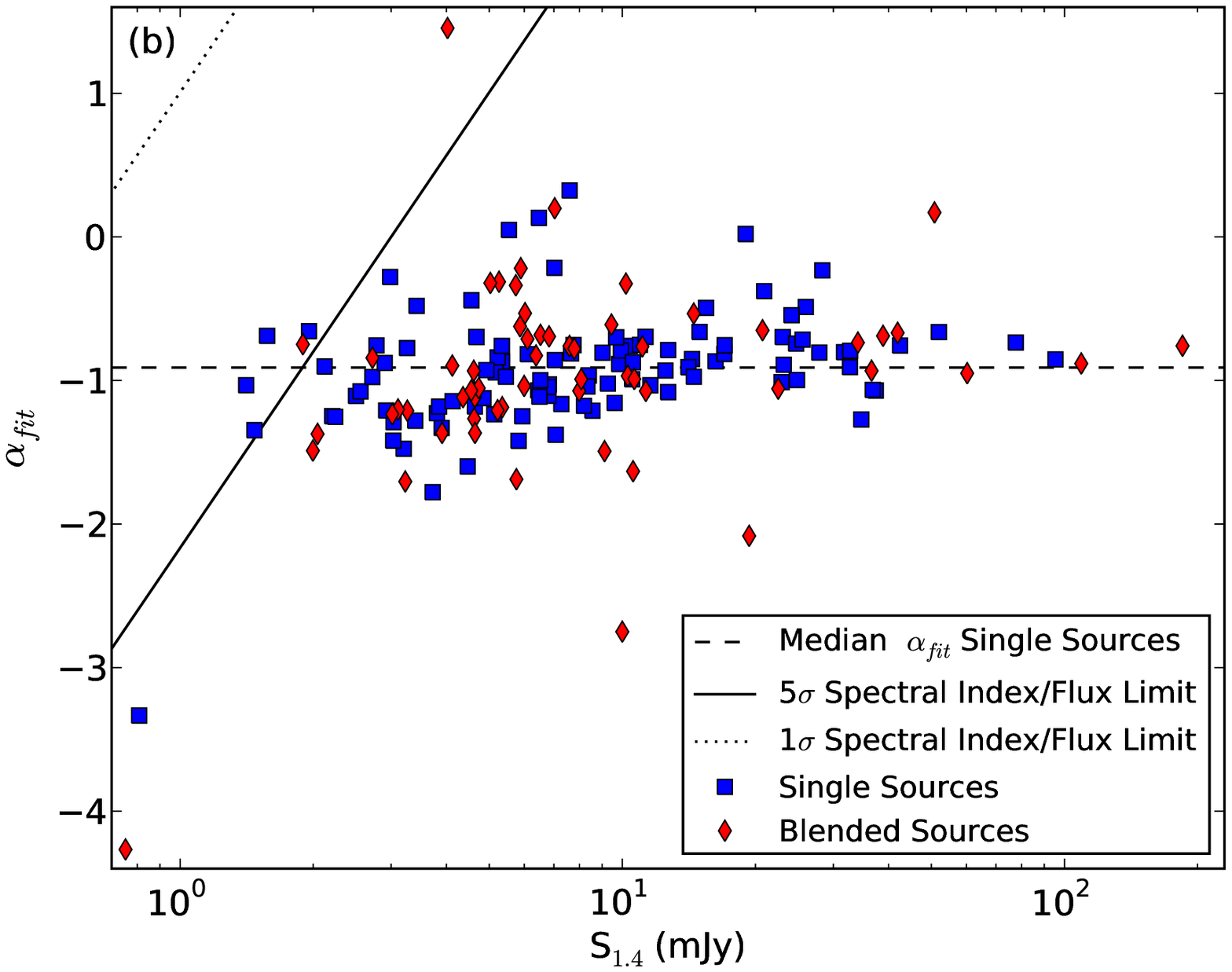}\includegraphics[scale=0.3]{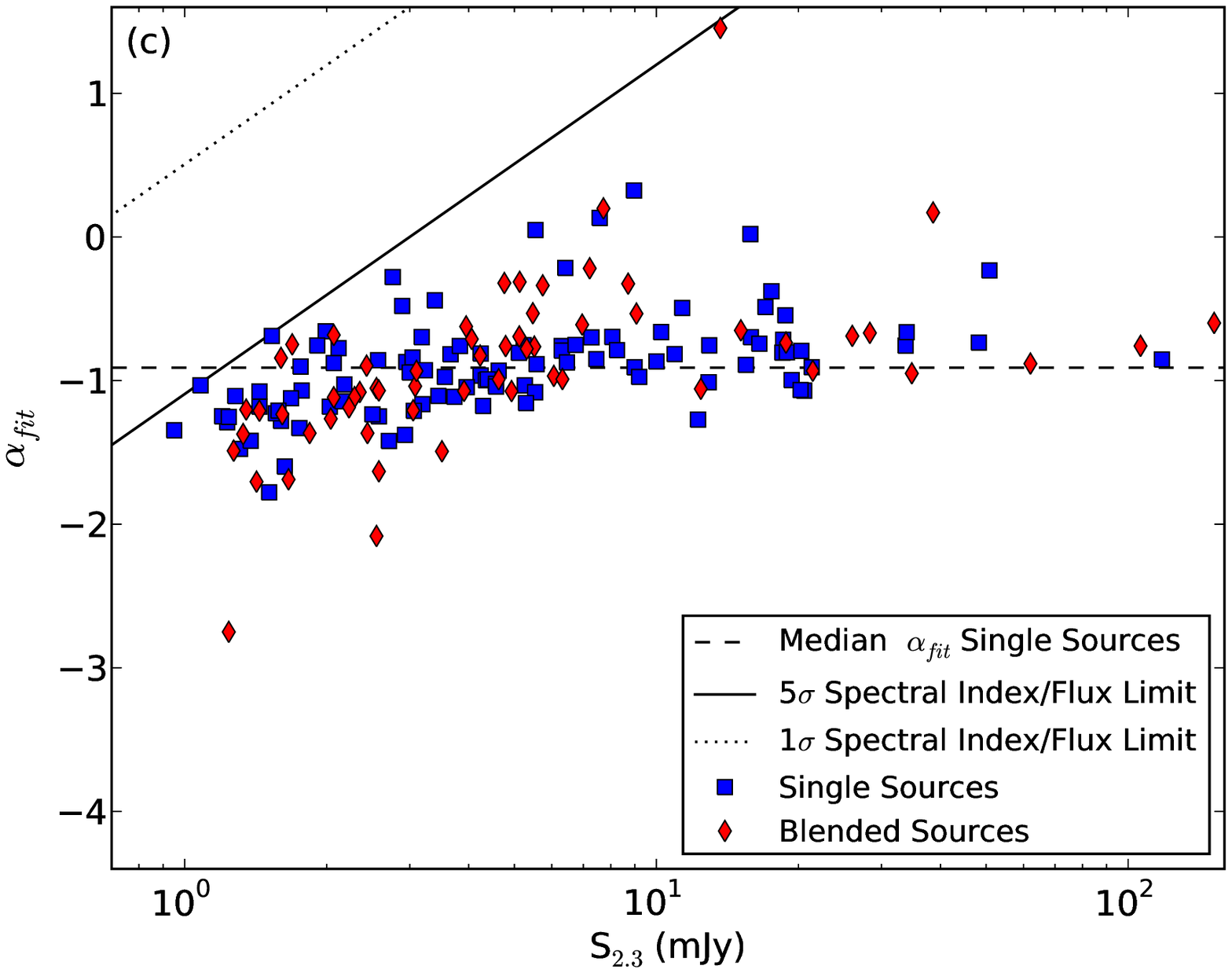}
\caption{(a) Spectral index $\alpha_{fit}$ versus 843\,MHz flux density, (b) 1.4\,GHz flux density, and (c) 2.3\,GHz flux density. The dotted black line indicates the $1\sigma$ spectral index limit, and the solid black line represents the $5\sigma$ spectral index limit for each flux density. The dashed black line is the median $\alpha_{fit}$ value for single sources only. 
\label{fig:alphaflux}}
\end{center}
\end{figure*} 
\section{The MOST ATLAS source Catalogue}
\label{sec:catalogue}
The total catalogue consists of 325 MOST sources, limited to sources above the 5$\sigma$ cutoff of 3\,mJy, of which 166 have an ATLAS 1.4 or 2.3\,GHz counterpart. The remaining 159 sources are outside the ATLAS survey area. In total, there are 310 ATLAS sources in our catalogue, with 205 classified as blended sources. The remaining 105 ATLAS sources correspond to single MOST sources. An extract of the catalogue in shown in Table~\ref{table:catalogue}, and the full version is available online. The MOST radio position is listed first, followed by the flux density and error at 843\,MHz. For MOST sources with a single ATLAS counterpart, the ATLAS source name (e.g. ATELAIS J002905.22$-$433403.9), and ID (e.g. S100) is listed; for blended sources the given ATLAS IDs correspond to all Z11 sources cross-matched to the MOST source, and the ATLAS source name corresponds to the first ATLAS ID listed. All ATLAS source names and IDs are from M08. The corrected convolved 1.4\,GHz and Z11 2.3\,GHz flux density measurements and associated errors are given, along with the three spectral indices (described in \S\ref{sec:specs} below).
 
\section{Results and Analysis}
\label{sec:results} 
\subsection{Flux Density Distribution}
\label{sec:fluxdist}
The flux density distributions (Figure~\ref{fig:fluxes}) do not show any major differences between the single, blended, and non-ATLAS sources. The location of the median of the entire catalogue is shown in Figure~\ref{fig:fluxes}a. The distributions appear consistent with other faint radio samples, such as the original ATLAS catalogues \citep{ray,Middelberg08}.

\begin{figure}
\begin{center}
\includegraphics[scale=0.4]{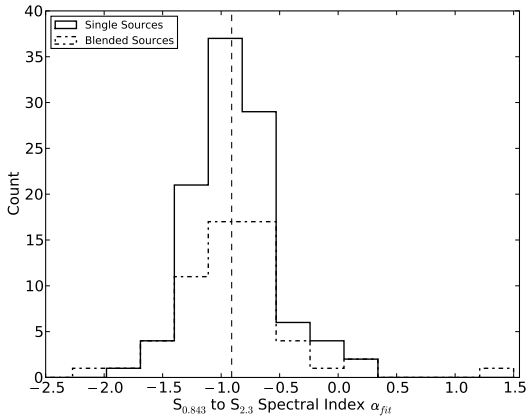}
\caption{Distribution of $\alpha_{fit}$ for our catalogue. 
\label{fig:alphahisto}}
\end{center}
\end{figure}

The Sydney University Molonglo Sky Survey \citep[SUMSS;][]{bock,SUMSS2} is an 843\,MHz survey with MOST, that covers the sky south of $\delta<-30^{\circ}$ with $|\textit{b}|>10^{\circ}$. SUMSS has similar resolution and sensitivity to the National Radio Astronomy Observatories (NRAO) Very Large Array (VLA) Sky Survey \citep[NVSS;][]{NVSS}. SUMSS contains 211063 radio sources, with an rms of $\sim1$\,mJy. Our observations probe sources fainter than SUMSS by a factor of $\sim2$. Of our catalogue, 178 sources are cross-matched to SUMSS sources. Only one source in SUMSS does not have a counterpart in our observations, as it is located directly on an artifact in our image. Our measured flux densities, positions, and radio differential source counts (see \S\ref{sec:srccnts} for details) are consistent with those of SUMSS, with only two outliers, likely due to intrinsic source variability. It is known that a few percent of mJy radio sources are variable on the timescale of years \citep{oort}. Attributing our outliers to variability is consistent with this work, as our data was taken over several years.

\subsection{Spectral Index Distributions and Properties} 
\label{sec:specs}
We have calculated spectral indices for all sources for which flux densities are available at two or three frequencies. $\alpha_{fit}$ is a three-point power-law fit, and $\alpha^{1.4}_{0.843}$ and $\alpha^{2.3}_{1.4}$ are two-point power-law fits for the respective frequencies. Distributions of these spectral indices are given in Figures~\ref{fig:alphaflux} and \ref{fig:alphahisto}. Spectral index $\alpha^{1.4}_{0.843}$ versus spectral index $\alpha^{2.3}_{1.4}$ is also shown in Figure~\ref{fig:specspec}, which indicates that most of our sources are steep-spectrum, with a small fraction of inverted, peaked, and flat spectrum objects, discussed further below.  A large proportion of steep-spectrum sources are seen because of the low radio frequency selection, and steep spectrum sources are brighter at lower frequencies. The median spectral index $\alpha_{fit}$ for the single sources is shown on each of the panels in Figure~\ref{fig:alphaflux}. Only single sources are included in this calculation. The spectral index limits are shown in Figure~\ref{fig:alphaflux}b,c arising from the limiting flux density of the least-sensitive frequency in the relevant two-point spectral index calculation. Although the three-point spectral index
\begin{landscape}
\begin{center}
\begin{deluxetable}{|ccccccccccccccc|}
\tablewidth{0pt}
\tabletypesize{\scriptsize}
\tablecaption{Extract from the 843\,MHz MOST ATLAS Catalogue \label{table:catalogue}}
\tablehead{
\colhead{MOST RA} & \colhead{MOST Dec} & \colhead{$S_{0.843}$} & \colhead{$\Delta\,S_{0.843}$} & \colhead{ATLAS Name} & \colhead{ATLAS ID/s} & \colhead{Type} & \colhead{$S_{1.4}$} & \colhead{$\Delta\,S_{1.4}$} & \colhead{$S_{2.3}$} &  \colhead{$\Delta\,S_{2.3}$} & \colhead{$\alpha_{fit}$} & \colhead{$\Delta\,\alpha_{fit}$} & \colhead{$\alpha^{1.4}_{0.843}$} & \colhead{$\alpha^{1.4}_{2.3}$}\\
 & & \colhead{\footnotesize mJy} & \colhead{\footnotesize mJy} & & &  & \colhead{\footnotesize mJy} & \colhead{\footnotesize mJy} & \colhead{\footnotesize mJy} &\colhead{\footnotesize mJy} & & & & \\
}
\startdata
0:28:45.438&$-$42:51:39.46&15.12&1.5&...&...&...&...&...&...&...&...&...&...&...\\
0:28:54.063&$-$43:12:18.30&6.80&1.8&...&...&...&...&...&...&...&...&...&...&...\\
0:28:56.980&$-$42:18:15.37&40.52&3.2&...&...&...&...&...&...&...&...&...&...&...\\
0:29:05.054&$-$43:34:07.05&11.34&1.3&ATELAIS J002905.22$-$433403.9&S749&p&9.01&1.16&5.11&0.34&$-$0.81&0.06&$-$0.45&$-$1.14\\
0:29:05.968&$-$44:42:00.59&14.18&1.6&...&...&...&...&...&...&...&...&...&...&...\\
0:29:09.167&$-$43:44:02.10&11.13&1.3&ATELAIS J002909.26$-$434356.3&S617&p&12.52&1.27&4.63&0.27&$-$0.93&0.09&0.23&$-$2.00\\
0:29:13.163&$-$44:52:23.32&5.33&1.4&...&...&...&...&...&...&...&...&...&...&...\\
0:29:15.437&$-$43:26:36.78&9.51&1.3&ATELAIS J002915.52$-$432638.3&S868&p&5.45&1.10&3.56&0.25&$-$0.97&0.03&$-$1.10&$-$0.86\\
0:29:21.471&$-$42:55:45.26&44.87&2.6&...&...&...&...&...&...&...&...&...&...&...\\
0:29:25.667&$-$44:08:25.68&15.37&1.8&ATELAIS J002925.66$-$440822.8&S293, S304&b&9.12&1.13&3.51&0.35&$-$1.49&0.06&$-$1.03&$-$1.92\\
0:29:27.845&$-$43:16:15.59&6.61&1.2&ATELAIS J002927.69$-$431614.4&S1014&p&3.85&1.16&2.03&0.18&$-$1.18&0.01&$-$1.06&$-$1.29\\
0:29:36.620&$-$42:25:41.20&64.07&8.4&...&...&...&...&...&...&...&...&...&...&...\\
0:29:37.206&$-$42:34:18.69&5.05&1.4&...&...&...&...&...&...&...&...&...&...&...\\
0:29:38.084&$-$44:23:21.78&42.38&2.7&ATELAIS J002939.19$-$442319.3&S100, S101.1&b&34.1&2.29&18.82&1.88&$-$0.74&0.06&$-$0.43&$-$1.20\\
0:29:43.820&$-$42:37:47.77&17.83&1.7&...&...&...&...&...&...&...&...&...&...&...\\
0:29:45.687&$-$43:21:50.03&24.18&1.6&ATELAIS J002945.64$-$432149.5&S943&p&16.25&1.41&10&0.52&$-$0.87&0.06&$-$0.78&$-$0.98\\
0:29:46.493&$-$43:15:57.34&41.67&2.4&ATELAIS J002946.52$-$431554.5&S1018&p&27.89&1.78&18.49&0.94&$-$0.81&0.05&$-$0.79&$-$0.83\\
0:29:47.622&$-$44:16:15.59&6.62&1.1&ATELAIS J002947.37$-$441607.0&S181&p&3.90&1.65&1.75&0.15&$-$1.33&0.01&$-$1.04&$-$1.62\\
0:29:50.398&$-$44:05:48.45&7.16&1.1&ATELAIS J002949.89$-$440541.4&S345, S342, S339&b&5.26&1.07&5.13&0.51&$-$0.31&0.03&$-$0.61&$-$0.05\\
0:29:51.452&$-$43:45:28.25&8.23&1.3&ATELAIS J002951.48$-$434528.0&S598&p&4.93&1.13&3.23&0.18&$-$0.93&0.02&$-$1.01&$-$0.85\\
0:29:53.062&$-$42:50:17.87&5.16&2.0&...&...&...&...&...&...&...&...&...&...&...\\
0:30:03.253&$-$42:18:22.97&20.81&8.4&...&...&...&...&...&...&...&...&...&...&...\\
0:30:04.749&$-$42:09:58.88&19.10&5.9&...&...&...&...&...&...&...&...&...&...&...\\
0:30:10.595&$-$44:09:12.06&6.49&1.1&ATELAIS J003010.82$-$440907.3&S288&p&7.60&0.72&8.96&0.46&0.32&0.04&0.31&0.33\\
0:30:17.439&$-$42:24:47.35&436.20&22.4&...&...&...&...&...&...&...&...&...&...&...\\
0:30:18.530&$-$45:26:42.05&12.79&2.5&...&...&...&...&...&...&...&...&...&...&...\\
0:30:18.793&$-$44:04:35.34&13.70&1.2&ATELAIS J003019.22$-$440438.3&S355&p&9.83&0.78&5.57&0.23&$-$0.89&0.08&$-$0.65&$-$1.14\\
0:30:20.948&$-$43:39:44.51&69.27&3.7&ATELAIS J003020.95$-$433942.8&S694&p&51.99&2.55&33.92&1.7&$-$0.66&0.04&$-$0.57&$-$0.86\\
0:30:21.663&$-$45:05:09.55&16.15&3.3&...&...&...&...&...&...&...&...&...&...&...\\
0:30:22.709&$-$42:37:02.29&14.25&2.4&...&...&...&...&...&...&...&...&...&...&...\\
0:30:27.017&$-$42:35:14.14&7.34&3.1&...&...&...&...&...&...&...&...&...&...&...\\
0:30:29.096&$-$42:13:50.96&18.19&6.2&...&...&...&...&...&...&...&...&...&...&...\\
0:30:34.835&$-$45:29:47.20&15.73&4.1&...&...&...&...&...&...&...&...&...&...&...\\
0:30:35.339&$-$44:37:11.21&3.69&1.1&ATELAIS J003035.77$-$443707.2&S18, S19&b&6.53&3.21&2.07&0.21&$-$0.68&0.01&1.13&$-$2.31\\
0:30:35.937&$-$43:23:39.75&8.95&3.1&ATELAIS J003035.03$-$432341.6&S926, S923, S930, S930.1&b&5.03&0.62&4.76&0.48&$-$0.32&0.02&$-$1.14&$-$0.11\\
0:30:38.957&$-$44:09:56.34&3.95&0.9&ATELAIS J003039.03$-$441000.0&S279&p&3.04&0.62&1.23&0.13&$-$1.29&0.01&$-$0.52&$-$1.82\\
0:30:39.015&$-$45:07:07.04&33.29&2.7&...&...&...&...&...&...&...&...&...&...&...\\
0:30:39.621&$-$44:42:01.21&28.43&1.9&ATELAIS J003039.68$-$444159.5&S7&p&26.00&3.44&17.02&0.95&$-$0.49&0.05&$-$0.18&$-$0.85\\
0:30:40.860&$-$43:23:40.55&11.17&2.4&ATELAIS J003042.10$-$432335.4&S923, S930, S930.1&b&6.02&0.64&5.47&0.55&$-$0.53&0.03&$-$1.22&$-$0.19\\
0:30:42.041&$-$43:18:42.39&4.52&1.0&ATELAIS J003041.88$-$431840.7&S987&p&3.26&0.61&2.12&0.13&$-$0.77&0.01&$-$0.64&$-$0.87\\
\enddata\\
\textsc{Notes.}-- $\alpha_{fit}$ is the spectral index fitted across the three flux density measurements at 0.843, 1.4 and 2.3\,GHz. Type (Column 7) refers to whether the source is a single point source (p) at all frequencies, or a blended source (b).\\
\end{deluxetable}
\end{center}
\end{landscape}
\noindent
 $\alpha_{\rm fit}$ is shown in the figure, the limit associated with the two-point spectral index calculation provides a clear indication of where we are selection-limited against particularly steep or flat spectrum sources. Distributions of $\alpha^{1.4}_{0.843}$, split into three flux density bins are also shown in Figure~\ref{fig:fluxbins}, analogous to Figure 8 of \citet{SUMSS2}. A flattening of the spectral index $\alpha^{1.4}_{0.843}$ is suggested by Figure~\ref{fig:fluxbins}; however this is primarily due to different numbers of sources in each flux density bin. This is further discussed in \S\ref{sec:specflux}.
\subsection{Radio spectral classifications}
\begin{figure}
\begin{center}
\includegraphics[scale=0.4]{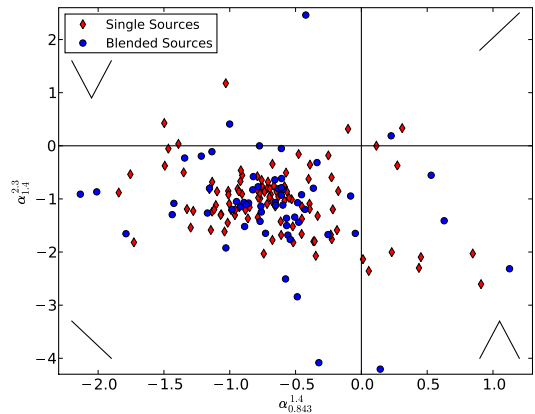}
\caption{Spectral index $\alpha^{2.3}_{1.4}$ versus $\alpha^{1.4}_{0.843}$ for MOST ATLAS sources. The black symbols in each outer corner of the plot represent the type of source in each quadrant of the plot, e.g. steep, inverted, peaked or upturn. 
\label{fig:specspec}}
\end{center}
\end{figure}

\begin{figure}
\begin{center}
\includegraphics[scale=0.4]{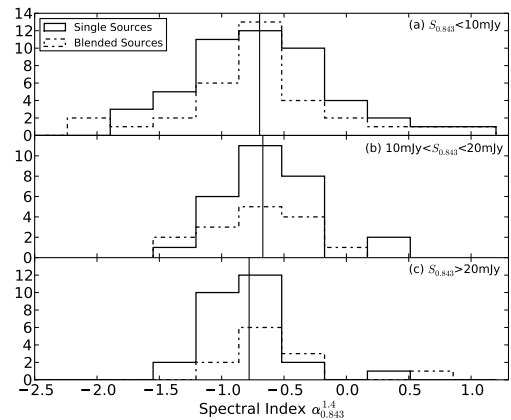}
\caption{Distribution of $\alpha^{1.4}_{0.843}$, in three flux density bins, 0-10\,mJy (a), 10-20\,mJy (b), and $>20$\,mJy (c). The solid vertical lines are the median spectral index $\alpha^{1.4}_{0.843}$ for each flux density bin for the single sources only.
\label{fig:fluxbins}}
\end{center}
\end{figure}

Figure~\ref{fig:specspec} shows our sample of sources, in a radio colour-colour diagram \citep{kesteven,at20gcat}, that we now consider in four classes:
\begin{enumerate}
\item Steep-spectrum objects, with a steep radio spectrum from 843\,MHz to 2.3\,GHz (the lower-left quadrant of Figure~\ref{fig:specspec}),
\item Peaked sources, where we see the radio spectrum turn over between 843\,MHz and 2.3\,GHz (the lower-right quadrant of Figure~\ref{fig:specspec}),
\item Inverted (or rising) sources, where the radio flux density increases with increasing frequency (the upper-right quadrant of Figure~\ref{fig:specspec}), and,
\item Upturn sources, that have an upturn in their radio spectrum (the upper-left quadrant of Figure~\ref{fig:specspec}).
\end{enumerate}
This distribution highlights the fact that our sample is dominated by steep-spectrum sources. The statistics are given in Table~\ref{table:class}, split into single and blended sources, noting that any statistics from the blended sources are not reliable. In comparison, the Australia Telescope 20 GHz Survey \citep[AT20G;][]{at20gcat}, a high-frequency-selected sample (20\,GHz), produced a colour-colour plot contained 3763 sources, with 14\% inverted (rising) spectrum sources, 57\% steep-spectrum objects, 21\% peaked spectrum sources, and 8\% sources with an upturn in their spectra. 

As indicated by Table~\ref{table:class}, 8\% of our MOST ATLAS sample appear to be possible Gigahertz Peaked Spectrum (GPS) sources (see \S\ref{sec:CSS} for a description). \citet{Review} suggest that $\sim10$\% of bright radio sources are GPS sources, whereas \citet{randall} found less than 1\% of their sample to be GPS sources. Although both these samples are not complete, it is interesting to note that at faint radio fluxes, the proportion of GPS sources appears to be similar to the bright sample of \citet{Review} rather than the unbiased bright sample of \citet{randall}. This sample of sources will be explored further in a future paper, \citet{randall1}.

Interestingly, there is one very steep spectrum single source, S1256, which has $\alpha^{1.4}_{0.843}=-2.92$, that is not detected in the 2.3\,GHz image. This object is discussed in more detail in \S\ref{sec:s1256}.

\begin{table}
\begin{center}
\caption{Spectral Classifications}
\label{table:class}
\begin{tabular}{|ccccc|} 
\hline
Type & Total & Point & Blended & Percentage\\
& Number & Sources & Sources & \\
\hline
Steep-Spectrum & 143 & 89 & 54 & 86\%\\
Inverted & 3 & 2 & 1 & 2\%\\
Peaked & 12 & 8 & 4 & 8\%\\
Upturn & 6 & 4 & 2 & 4\%\\
\hline
\end{tabular}
\textsc{Notes.}-- The percentages of each source are only for the single or point sources. \\
\end{center}
\end{table}

\subsection{Spectral index as a function of flux density}
\label{sec:specflux}
\begin{figure*}
\begin{center}
\includegraphics[scale=0.4]{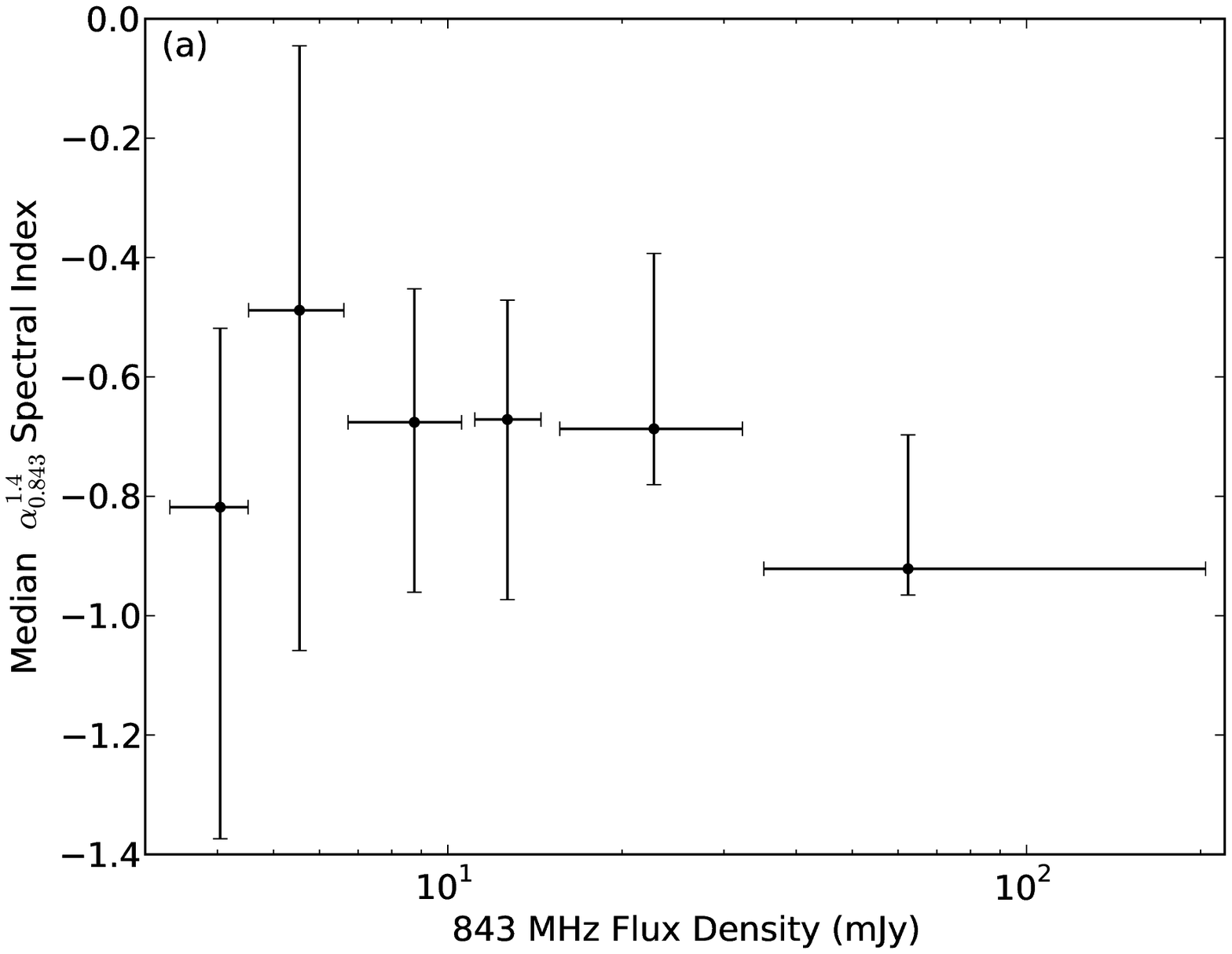}\includegraphics[scale=0.4]{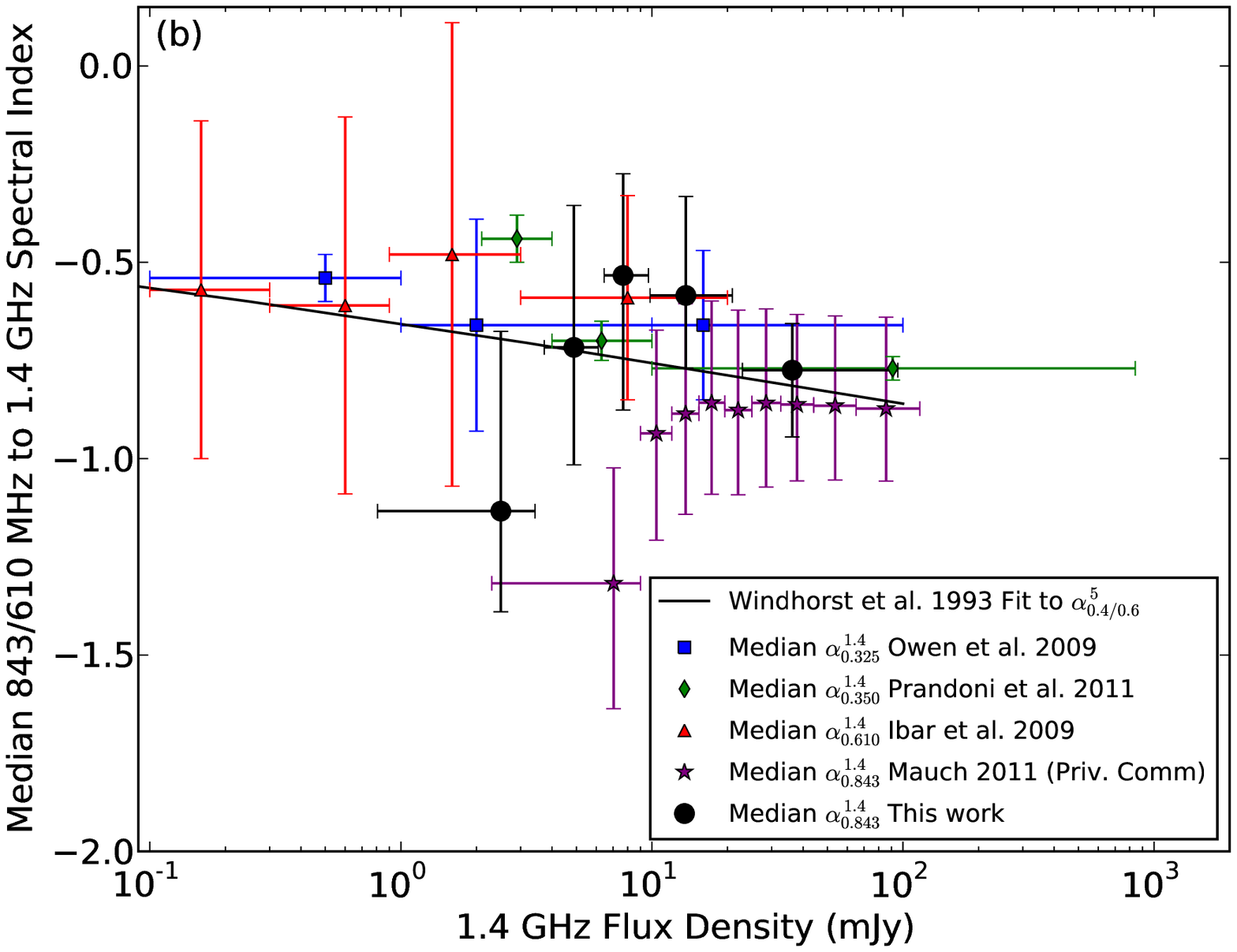}
\caption{(a) Median spectral index $\alpha^{1.4}_{0.843}$ for our sample with 843\,MHz flux density. (b) Median spectral index $\alpha^{1.4}_{0.843}$ for our sample, compared to median spectral indices from the literature. The Windhorst et al. 1993 solid line indicates a fit to the median spectral index $\alpha^{5}_{0.4/0.6}$ for their data. The other values are taken from tabulated results in Ibar et al., 2009, Owen et al., 2009, Prandoni et al., 2011, and Mauch et al. 2011 (Priv. Comm). The horizontal error bars represent the flux density bin for each median point; the vertical error bars are the 25$^{th}$ and 75$^{th}$ percentiles.
\label{fig:medians}}
\end{center}
\end{figure*}

We have investigated the properties of our sample with flux density. Figure~\ref{fig:medians} shows median spectral index as a function of flux density, both for our sample (Figure~\ref{fig:medians}a), and for a compilation of samples from the literature along with ours (Figure~\ref{fig:medians}b). While the uncertainties are large, there is evidence for a mild trend toward a flatter spectral indices with decreasing flux density (Figure~\ref{fig:medians}a). Figure~\ref{fig:medians}b shows the comparison of our median spectral indices with \citet{windhorst,prandoni4,ibar}, and \citet{owen1}. Our data is in general consistent with these previous results, and so we cannot rule out the possibility that there is a flattening of the median spectral index with decreasing flux density. We note that the very steep median spectral index in the faintest flux density bin of many surveys is a consequence of the flux density limits preventing the detection of flatter spectral indices close to the survey limits. 

\begin{table}
\addtolength{\tabcolsep}{-3pt}
\fontsize{8}{8}\selectfont
\begin{center}
\caption{Median Spectral Index $\alpha^{1.4}_{0.843}$ Statistics as a function of flux density}
\label{table:medians}
\begin{tabular}{|cccccc|} \hline
Range in $S_{843}$ & Median & Mean Flux & Number & 25$^{th}$ & 75$^{th}$\\
or $S_{1.4}$ (mJy) &  $\alpha^{1.4}_{0.843}$ & (mJy) & of sources & perc. & perc.\\
\hline
$S_{1.4}$ & & & & \\
\hline
0.81-3.7 & $-$1.134 & 2.50 & 21 & $-$1.390 & $-$0.676\\
3.7-6.4 & $-$0.717 & 4.89 & 21 & $-$1.016 & $-$0.355\\
6.4-9.8 & $-$0.533 & 7.68 & 22 & $-$0.876 & $-$0.275\\
13.5-26 & $-$0.584 & 13.65 & 21 & $-$0.769 & $-$0.332\\
26-95.4 & $-$0.775 & 36.21 & 20 & $-$0.945 & $-$0.656\\
\hline
$S_{0.843}$ & & & & \\
\hline
3.3-4.52 & $-$0.818	& 4.04 & 18 & $-$1.273 & $-$0.825\\
4.52-6.7 & $-$0.488	& 5.54 & 18 & $-$1.203 & $-$0.733\\
6.7-11 & $-$0.676 & 8.76 &17 & $-$1.210 & $-$0.817\\
11-15 & $-$0.671 & 12.68 & 18	 & $-$1.045 & $-$0.795\\
15-35 & $-$0.687 & 22.71 & 18 & $-$0.872 & $-$0.575\\
35-204 & $-$0.921 & 62.42 & 16 & $-$0.930 & $-$0.741\\
\hline
\end{tabular}
\textsc{Notes.}-- Errors on the median spectral indices are calculated from the 25th and 75th percentile. The mean flux densities are calculated as the mean of the flux densities of all the sources in each bin.\\
\end{center}
\end{table}

\subsection{Radio Source Counts}
\label{sec:srccnts}
\begin{figure}
\begin{center}
\includegraphics[scale=0.45]{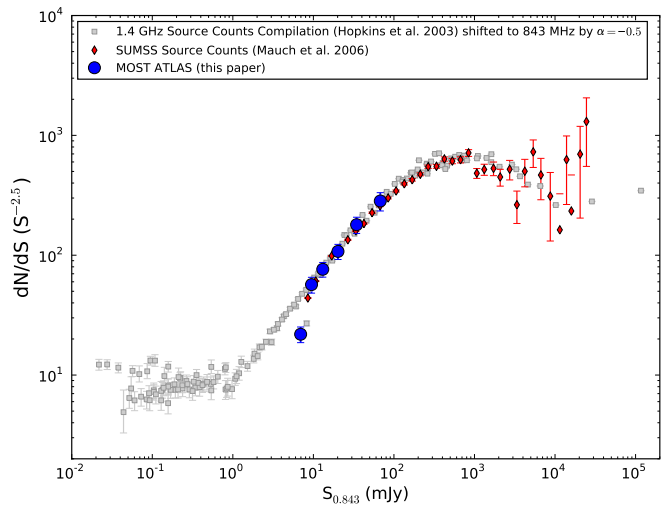}
\caption{The differential source counts for our data (circles), SUMSS (diamonds), and a 1.4\,GHz compilation of source counts (grey squares) shifted to 843\,MHz assuming a spectral index of $\alpha=-0.5$, from Hopkins et al. 2003 as our reference sample.
\label{fig:sourcecounts}}
\end{center}
\end{figure}

The differential radio source counts for our 843\,MHz data are presented in Table~\ref{table:mostsc}. A weighting factor has been applied to the source counts to correct for incompleteness due to the noise level increasing at the edges of the field \citep[see][]{srccnts}. We do not apply a resolution correction, as the MOST has a large beam, and has a high sensitivity to extended diffuse radio emission \citep{bock,SUMSS2}. The differential source counts are shown in Figure~\ref{fig:sourcecounts}, with the SUMSS \citep{SUMSS2} differential source counts, and a compilation of 1.4\,GHz differential source counts \citep{phoenix} as a reference sample. Our data probes the 843\,MHz source counts a factor of 2 fainter than SUMSS, but we still clearly underestimate the counts in our faintest flux density bin, due to incompleteness. 

The reference differential source count sample at 1.4\,GHz from \citet{phoenix} has been shifted to 843\,MHz assuming an average spectral index value. We have determined that $\alpha=-0.5$ is the appropriate value to use for the average spectral index, using the Kellermann correction, discussed in further detail here. 
\subsubsection{The Kellermann Correction}
\label{sec:kell}
\citet{kellermann} noted that any observed spectral index distribution is not independent of the observing frequency. The correction accounts for the relationship between observed spectral index distributions at different frequencies. At higher radio frequencies, we tend to see more flat spectrum objects as proportionally more sources are above the flux limit. In contrast, as we move to lower radio frequencies, we tend to see more steep spectrum objects because they are brighter at the lower radio frequencies. The correction stated in the Appendix of \citet{kellermann} describes the offset between mean values of spectral indices at different frequencies. We use Equation A5 of \citet{kellermann}, where we assume we have two distributions of spectral indices ($P(\alpha)$ and $Q(\alpha)$), at two different frequencies, $\nu_{1}$ and $\nu_{2}$ respectively. Also required is $x$, the slope of the number-flux power-law $N(>S)=k\,S^{x}$ \citep{longair}, where $k$ is a constant scaling factor, and $N$ represents the number of sources above a given flux density, $S$.
\begin{equation}
Q(\alpha)=A(\frac{\nu_{2}}{\nu_{1}})^{-\alpha\,x}P(\alpha),
\label{eq:kell}
\end{equation}
\begin{equation}
\int^{\infty}_{-\infty}Q(\alpha)d\alpha=1.
\label{eq:kell1}
\end{equation}
$P(\alpha)$ is the known distribution of spectral indices at one frequency and $Q(\alpha)$ is the distribution of spectral indices we wish to infer, given by Equation~\ref{eq:kell}. Following \citet{kellermann} we assume the special case where these two distributions are Gaussian (a reasonable assumption given the shapes of the distributions in Figures~\ref{fig:alphahisto} and \ref{fig:fluxbins}), and have the same dispersion $\sigma$, and A is chosen such that Equation A6 of \citet{kellermann} (Equation~\ref{eq:kell1}) is satisfied. This special case scenario results in the mean value of the inferred spectral index distribution ($Q(\alpha)$) being shifted by a factor of $x\sigma^{2}$ln$(\nu_{1}/\nu_{2})$. For our sample, to shift the distribution of the 1.4\,GHz source count compilation to 843\,MHz, we used $\alpha^{1.4}_{0.843}$ to determine the value of the Kellermann correction for only the single sources. The dispersion was found to be $\sigma=0.56$ and $x=-1.60$, resulting in a Kellermann correction of 0.25. This shift in the mean of the spectral index distribution from 843\,MHz to 1.4\,GHz is enough to account for the use of the $\alpha\sim-0.5$ to shift the 1.4\,GHz source counts to the 843\,MHz source counts, instead of the observed mean spectral index of our sample $\alpha=-0.71$.

\begin{table}
\begin{center}
\caption{Source Counts from the ATLAS ELAIS MOST Observations}
\label{table:mostsc}
\begin{tabular}{|ccccc|} \hline
Range in $S_{843}$ & \textit{N} & \textit{$N_{eff}$} & $\langle\,S_{843}\,\rangle$ & $(dN/dS)/S^{-2.5}$\\
(mJy) & & & (mJy) & (Jy$^{1.5}$sr$^{-1}$)\\
\hline
3.31-5.81 & 45 & 21.58 & 4.38 & 4.29\\
5.81-8.31 & 45 & 34.87 & 6.95 & 21.90\\
8.31-10.81 44 & 41.63 & 9.48 & 56.83\\
10.81-15.81 & 51 & 50.02 & 13.07 & 76.29\\
15.81-25.81 & 48 & 47.54 & 20.20 & 107.61\\
25.81-45.81 & 42 & 42 & 34.38 & 179.70\\
45.81-100 & 33 & 33 & 67.68 & 283.27\\
\hline
\end{tabular}\\
$N_{eff}$ is the effective number of sources in each bin, after the weighting factor has been applied. The mean flux density given is the geometric mean of the bin limits.
\end{center}
\end{table}

\section{Faint CSS and GPS candidates sources in ATLAS}
\label{sec:props} 
\subsection{Compact Steep Spectrum and Gigahertz Peaked Spectrum Sources}
\label{sec:CSS}
Gigahertz Peaked Spectrum (GPS) and Compact Steep Spectrum (CSS) sources \citep{young, cfanti, morganti} are a class of compact, powerful radio sources, suggested to be the beginning of the evolutionary path for large-scale radio sources. The proposed evolutionary path has GPS sources evolving into CSS sources, which then gradually evolve into Fanaroff-Riley Type I and II galaxies \citep{fr}, depending on their initial luminosity. These sources offer an ideal resource to investigate galaxy evolution and formation, as well as AGN feedback, as they are young AGN, but also have star formation occurring due to interactions and mergers \citep{Review,labiano,morganti}. A more detailed discussion of their properties is given in \citet{randall}. If bright CSS sources do evolve into FRI/II's \citep{fr}, then some strong evolution must occur for this to happen \citep{evolution}, as we do not see this high percentage of CSS and GPS sources in the local Universe. It is also possible that the supply mechanism of the energy powering these objects could cut off, leaving only diffuse emission and a steep spectrum core. This would result in faint objects whose radio lobes have ceased to expand or a prematurely dying radio source \citep{rfanti}, explaining the lack of large numbers of these objects nearby. 

Bright CSS and GPS sources are common ($\approx30\%$ and $10\%$ respectively) in radio surveys, but few faint samples exist. Previous surveys include \citet[][and references therein;]{snell00,tschager,kunert07,fanti}, which catalogue CSS and GPS sources down to $\sim20$\,mJy. If these objects are as prevalent at faint flux densities, a whole population of these objects remains mostly unknown \citep{faintgps,faint}. Here we present an initial complete sample of faint CSS and GPS sources from ATLAS, to further our understanding of their role in galaxy formation and evolution. Understanding their properties across different wavelength regimes is important, as is studying samples across a wide range of flux densities, as this will help to build a complete picture of their properties and nature. We present here a brief overview of the selection and properties of the faint sample of candidate CSS and GPS sources selected from ATLAS. 
\subsection{A new candidate sample of faint CSS sources}
\label{sec:newfaint}
An initial sample of faint CSS sources has been selected from ATLAS (\S\ref{sec:atlas}), based upon spectral index information, and angular size. The initial sample is drawn from both ATLAS fields, CDFS and ELAIS, and only includes the 1.4 and 2.3\,GHz flux density measurements. CDFS was not observed by MOST, and we do not consider the ELAIS MOST observations in this discussion. 

\begin{table*}
\begin{center}
\caption{\textsc{Summary of selection criteria and number of sources remaining}
\label{table:selection}}
\begin{tabular}[t]{|cccc|} 
\hline
 Field & & ELAIS & CDFS\\
 \hline
Criterion & Selection & Sources & Sources\\
Number & Criterion & Remaining & Remaining\\
\hline
1 & Unresolved in component catalogue & 681 & 271\\
2 & Not part of a radio double/triple/complex object or classified as a sidelobe & 638 & 247\\
3 & Apply spectral index cut of $\alpha^{2.3}_{1.4}<-0.95$ for sample selection & 78 & 8\\
3a & Sources with spectral index between $-0.9>\alpha^{2.3}_{1.4}>-0.95$ selected as supplementary sample & 4 & 1\\
\hline
\end{tabular}\\
\textsc{Notes.}-- Removal of radio doubles/triples/complex sources was done by cross-matching both the component, and source catalogues from CDFS and ELAIS, as the source catalogue lists what components comprise each source.
\end{center}
\end{table*}

The selection criteria listed in Table~\ref{table:selection} were utilized to first select a sample of unresolved, single objects from both the ELAIS and CDFS fields, before applying a spectral index cut. The number of sources remaining in each field after each criterion is also listed. We have chosen a primary spectral index cut of $\alpha^{2.3}_{1.4}<-0.95$ with an additional supplementary set extending to $\alpha^{2.3}_{1.4}<-0.9$, as the errors on the spectral indices are typically $\sim0.2$. These values were chosen to avoid selecting any star forming galaxies (with typical synchrotron spectral indices of $\alpha^{2.3}_{1.4}\sim-0.7$), as at these low flux densities, star forming galaxies are a dominant part of the population \citep{windhorst85,windhorst,phoenix,seymour}. In the ELAIS catalogue, there are 576 sources with a single 2.3\,GHz match (blended sources are not included in this analysis) out of 1276, and 399 of 726 in the CDFS. The primary reason for only half the sources having a 2.3\,GHz counterpart is the lack of equivalently deep 2.3\,GHz data across the two fields. The 2.3\,GHz data has lower resolution ($33''\times20''$ and $54''\times21''$ for ELAIS and CDFS respectively), and does not have the same sensitivity as the 1.4\,GHz data for both ELAIS and CDFS. The rms of the CDFS 2.3\,GHz data is $\sim3$ times higher than the ELAIS 2.3\,GHz data. The initial selection criterion of being an unresolved component also removes 63\% of the sources from CDFS, but only 47\% from ELAIS. The percentage of sources remaining after the initial selection is small, with 86 sources and 5 supplementary sources, from a possible 975 ($\sim10$\%). 

\begin{figure*}
\begin{center}
\includegraphics[scale=0.4]{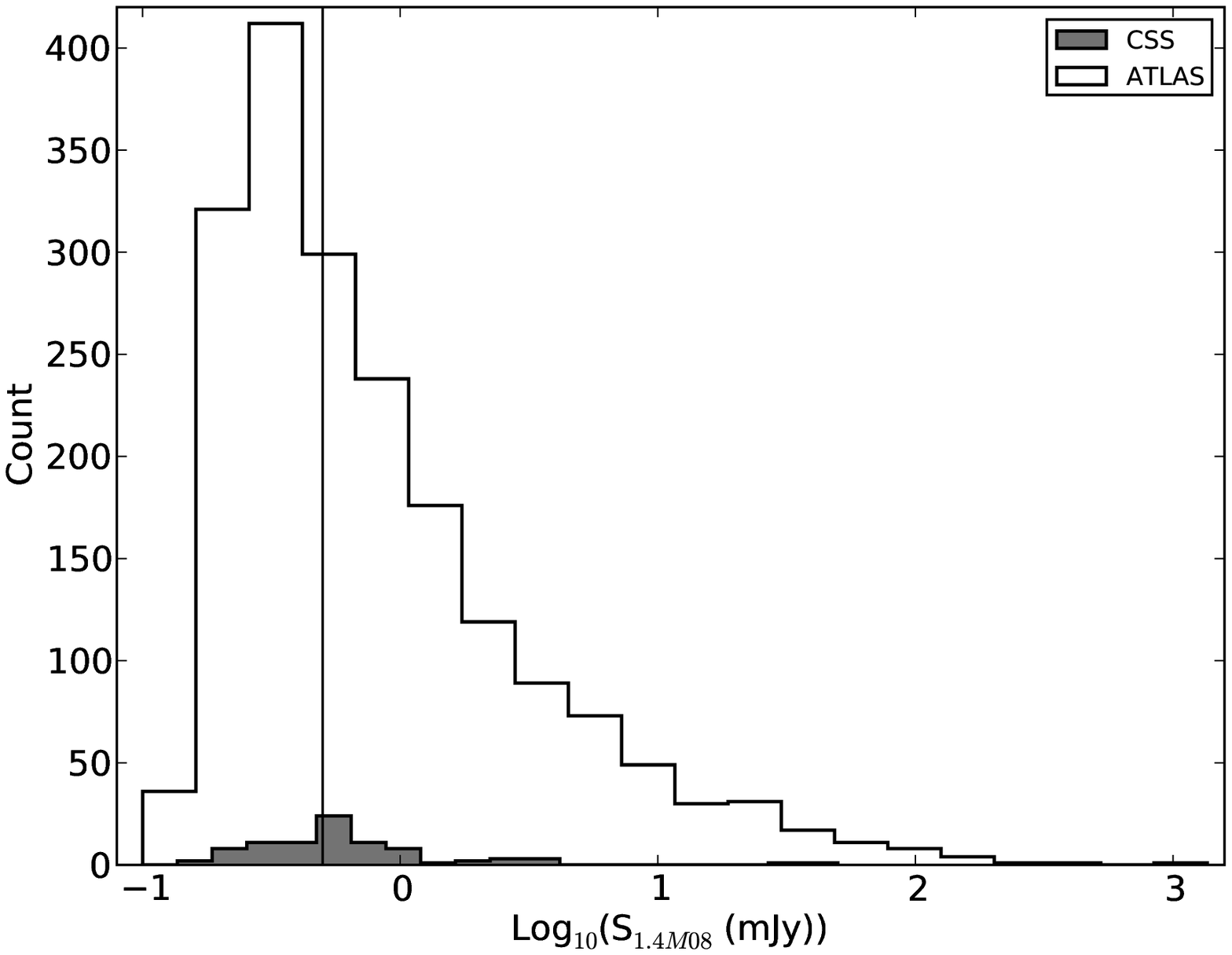}\includegraphics[scale=0.4]{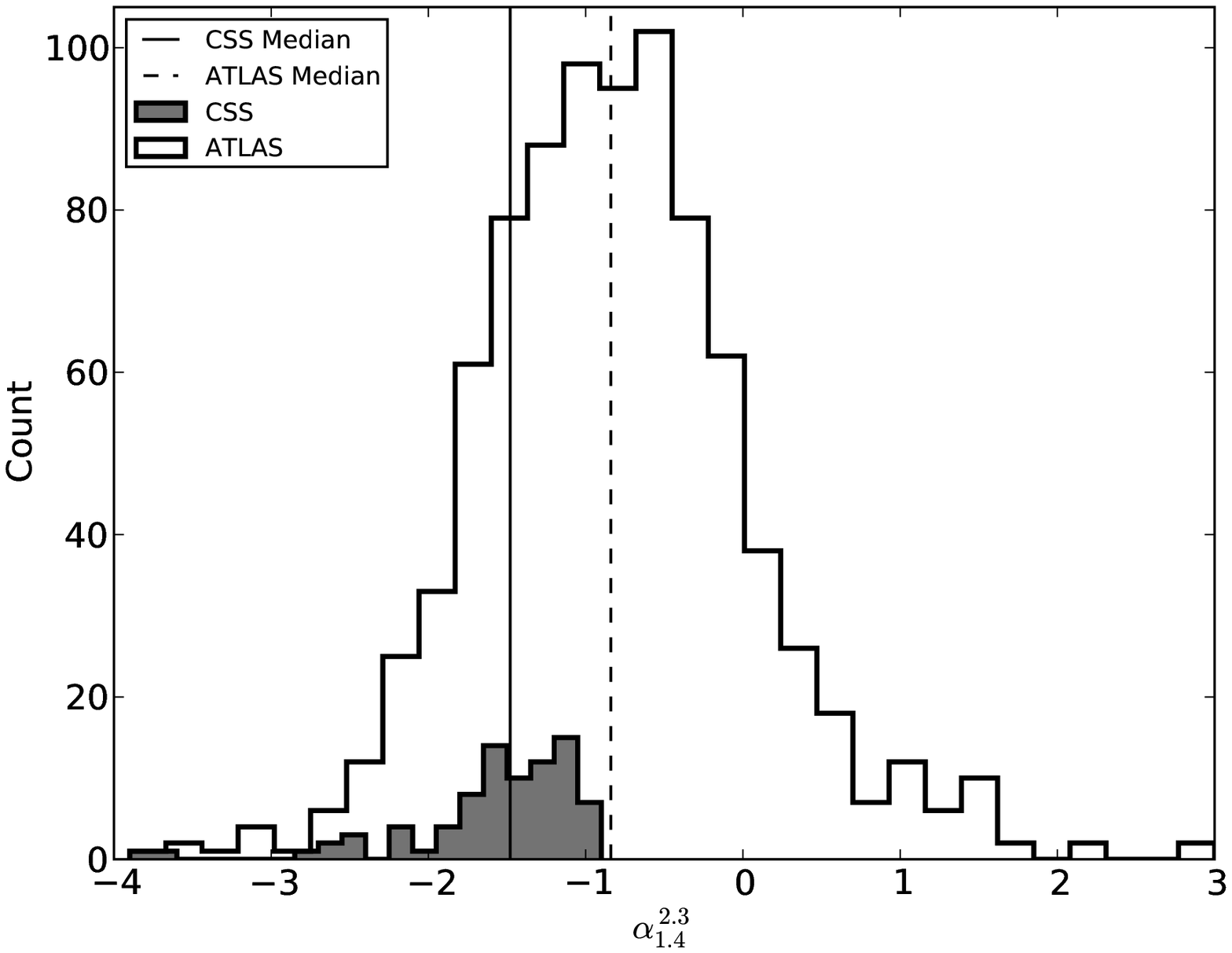}
\caption{(a) The 1.4\,GHz M08/N06 flux density distribution for the initial CSS sample, for the full and supplementary sets, shown as the grey filled distribution. The entire ATLAS catalogue is shown in the solid empty distribution. The solid vertical line represents the median of both distributions. (b) The solid grey filled histogram indicates the spectral index $\alpha^{1.4}_{2.3}$ distribution for the initial CSS sample, and the solid empty distribution is the entire ATLAS catalogue spectral index distribution. The vertical solid and dashed lines indicate the median of each distribution respectively.
\label{fig:fluxdist}}
\end{center}
\end{figure*}

Of the 86 sources, only 5 are cross-matched to 843\,MHz sources, due to the relatively high flux limit of the 843\,MHz data. The relevant information from these sources is added into the sample catalogue, but is not used for any subsequent analysis below. In considering the spectral index selection, described in Table~\ref{table:selection}, for each field, we note that the CDFS 1.4\,GHz flux densities were taken for the sources as observed, rather than after convolution to match the 2.3\,GHz resolution. This step has been omitted from the current analysis due to the small numbers of sources involved, and while the spectral index estimates for these few sources may be somewhat less robust than those in the ELAIS field, this does not have a significant impact on any of the results presented here.

In the ELAIS field, we use the spectral indices measured between 2.3\,GHz and the Z11 1.4\,GHz flux densities (measured with the same resolution as the 2.3\,GHz image). If the source does not have a reliable convolved measured flux density, we use the M08 flux (this is only necessary for four sources). To robustly constrain the radio spectra of these sources, we need to explore both higher, and lower frequency data to confirm that our sources are truly steep-spectrum or peaked spectrum objects. We show the basic statistics of our sample in Table~\ref{table:means}, and discuss initial results in \S\ref{sec:discussion}. The 1.4\,GHz flux density and spectral index distributions are presented in Figure~\ref{fig:fluxdist}, with the median flux density and spectral index included as the solid black lines. The spectral index distribution of the entire ATLAS catalogue is also shown in Figure~\ref{fig:fluxdist}b to provide context for the initial CSS sample distribution. Note that the steep spectrum sources not identified as CSS are those that were rejected by criteria 1 or 2 in Table~\ref{table:selection}. The median of the CSS sample is $\alpha^{2.3}_{1.4}=-1.48$, compared to $\alpha^{2.3}_{1.4}=-0.84$ for the entire ATLAS catalogue. 

\begin{figure*}
\begin{center}
\includegraphics[scale=0.4]{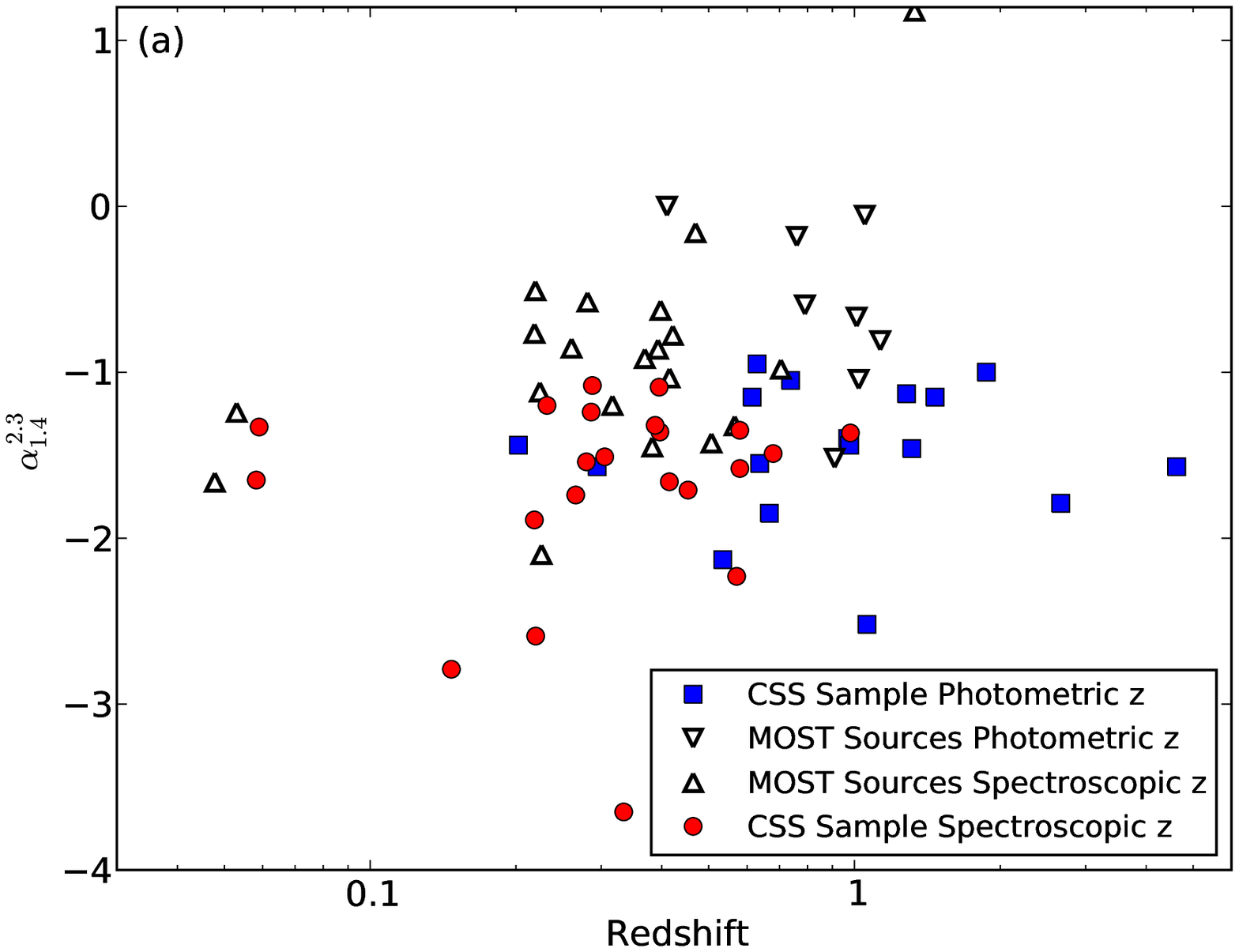}\includegraphics[scale=0.4]{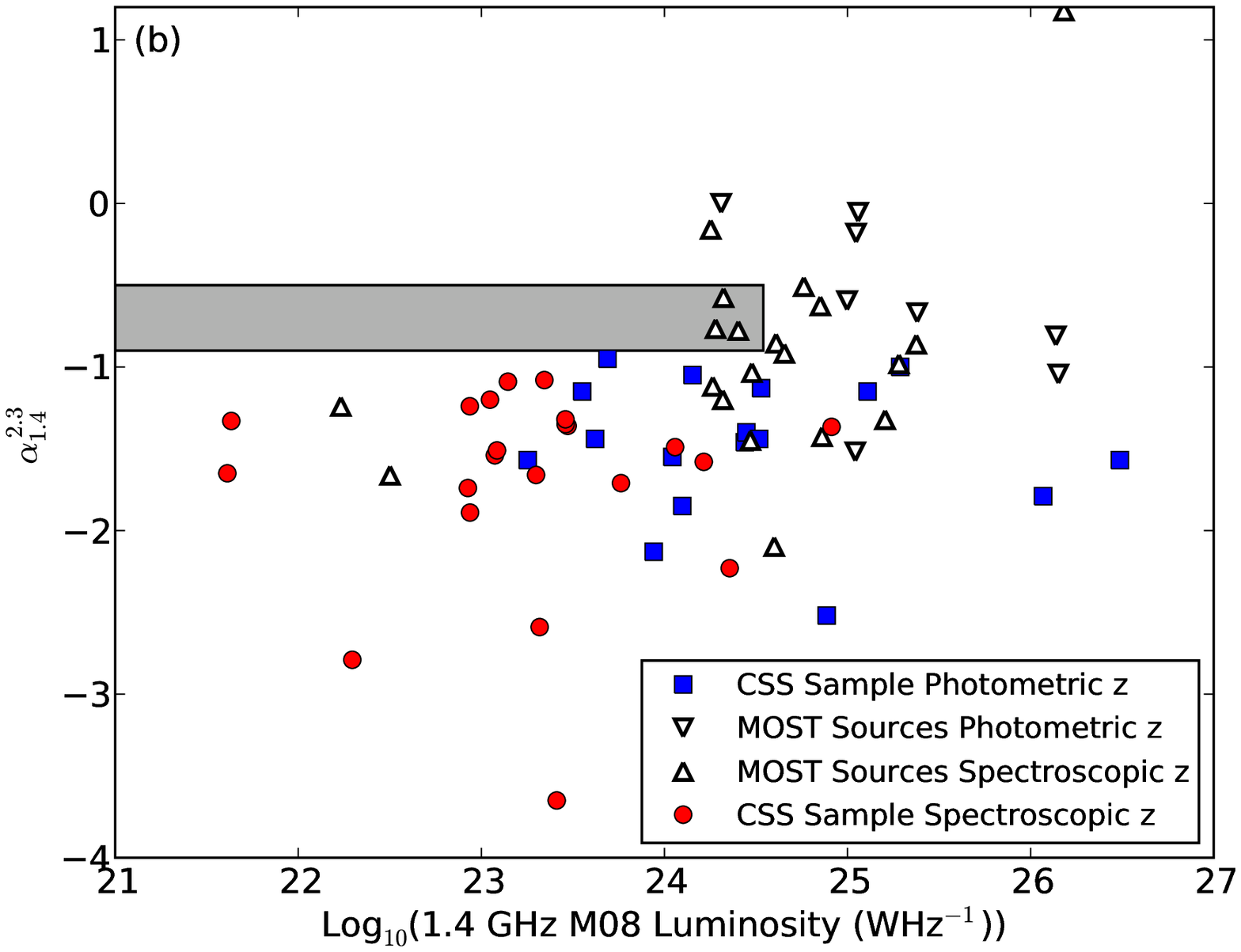}
\caption{(a) Redshift versus spectral index $\alpha^{2.3}_{1.4}$ for those sources of our MOST ATLAS sample, and initial CSS sample with spectroscopic and photometric redshifts. (b) Luminosity versus $\alpha^{2.3}_{1.4}$ for our samples at 1.4\,GHz. The pale grey box is the area where SFGs would dominate the population on the basis of having a synchrotron spectral index, and a radio luminosity consistent with a star formation rate up to $\sim1000\,M_{\odot}\,$yr$^{-1}$. Any sources in the remaining area are AGN.
\label{fig:lumalpha}}
\end{center}
\end{figure*}

\begin{table*}
\begin{center}
\caption{\textsc{Properties of our faint initial CSS Sample}
\label{table:means}}
\begin{tabular}{|ccccccccc|} \hline
& Spectral Index & Spectroscopic Redshift & Photometric Redshift & R magnitude & 1.4\,GHz flux density & 2.3\,GHz flux density\\
\hline
Mean & $-1.58$ & 0.37 & 1.21 & 19.03 & 1.68 & 1.08\\
Median & $-1.48$ & 0.32 & 0.97 & 18.8 & 0.52 & 0.36\\
\hline
\end{tabular}\\
\end{center}
\end{table*}

\section{Discussion}
\label{sec:discussion}
\subsection{Redshift and Luminosity Properties}
\label{sec:lums}
ATLAS has $\sim$700 spectroscopic \citep{mao}, and $\sim$700 photometric \citep{rowan} redshifts for radio sources in the CDFS and ELAIS-S1 fields. The sources chosen for the spectroscopic targets were primarily the optically bright sources (80\% complete at R magnitude $\sim$20), that are mainly low redshift sources, although some quasars at higher redshift are also selected. For the 105 single sources in the MOST ATLAS catalogue, we found 20 spectroscopic redshifts and 9 photometric redshifts (no analysis is done on the blended sources). In the initial CSS sample of 91, 22 spectroscopic and 17 photometric redshifts were found. In Figure~\ref{fig:lumalpha} spectral index against redshift and 1.4\,GHz rest-frame luminosity are shown, for both the MOST ATLAS catalogue and the initial CSS sample. The grey box in Figure~\ref{fig:lumalpha}b indicates where SFGs would most likely be located, defined by the dividing line between AGN and SF luminosity of  L$_{1.4}=\sim10^{24.5}$\,WHz$^{-1}$, and a spectral index between $-0.9<\alpha^{2.3}_{1.4}<-0.5$, typical for SFGs \citep{afonso}. This is a conservative upper limit on the star formation rate, assuming that the star formation rate implied from the radio emission is very high. The single inverted spectrum source with a spectroscopic redshift of $z=1.33$ is classified spectroscopically as a quasar, with broad emission lines. It has an unusual radio spectrum that appears to contain an upturn, with $\alpha^{1.4}_{0.843}=-1.03$, and $\alpha^{2.3}_{1.4}=1.18$, one of only 6 MOST ATLAS sources with an upturning radio spectrum. 
\begin{figure}
\begin{center}
\includegraphics[scale=0.4]{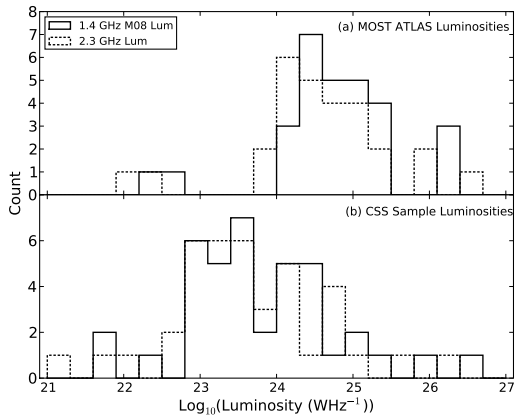}
\caption{(a) Distribution of luminosities for our MOST ATLAS sources and (b) initial CSS sample, at 1.4\,GHz and 2.3\,GHz. 
\label{fig:lums}}
\end{center}
\end{figure}

The rest-frame 843\,MHz, 1.4\,GHz, and 2.3\,GHz luminosities were calculated for each source in the MOST ATLAS catalogue with a redshift (spectroscopic or photometric), and the 1.4 and 2.3\,GHz luminosities are shown in Figure~\ref{fig:lums}a. The rest-frame 1.4 and 2.3\,GHz luminosities were also calculated for the CSS sample, and are shown in Figure~\ref{fig:lums}b. The distribution of luminosities, in conjunction with their spectral index, is consistent with these sources being AGN, as expected. Although it appears in Figures~\ref{fig:lumalpha}b and \ref{fig:lums}b that the CSS sample is on average a lower-luminosity sample, this is due to the flux limit of the 843\,MHz MOST data, that preferentially selects the higher-luminosity objects.  However, as we only find 4 of these MOST ATLAS sources in our CSS candidate sample, we are selecting a low-luminosity sample of young AGN that is ideal for comparison to our bright sample \citep{randall}. 

\begin{figure*}
\begin{center}
\includegraphics[scale=0.8]{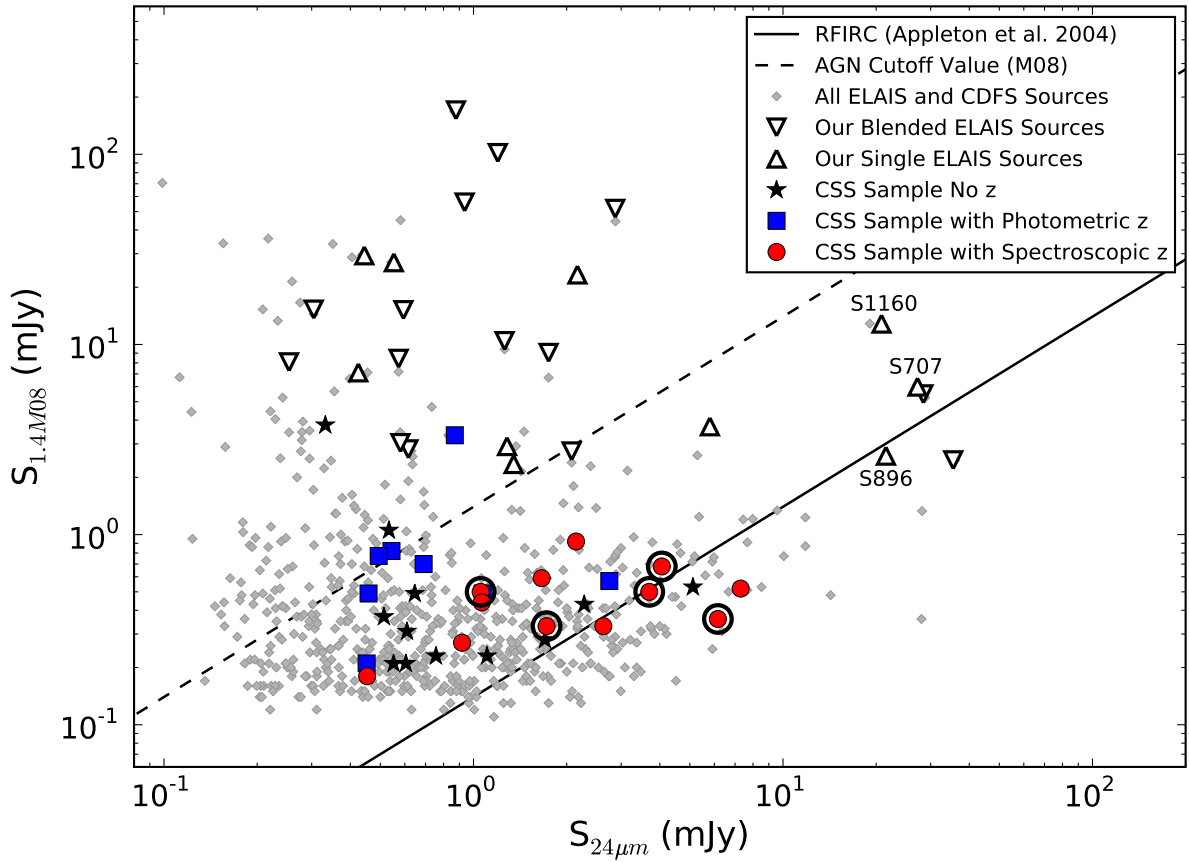}
\caption{The radio-far infrared correlation for our sample, and the ATLAS data for CDFS and ELAIS. The N06 and M08 1.4\,GHz fluxes are used for all sources for consistency. The initial sample of faint CSS sources are also shown, split into candidates with and without redshifts. Objects with a thick black circle around the data point are from the CSS sample and are those spectroscopically classified as SFGs. 
\label{fig:rfir}}
\end{center}
\end{figure*}

The MOST ATLAS and initial CSS sample sources with a spectroscopic redshift have also been spectroscopically classified as AGN or SFGs \citep{mao}. Of the 22 MOST ATLAS sources with spectroscopic redshifts, 15 sources were classified as AGN, four were classified as SFGs and the remaining two sources were unclassified. Three of the SFG sources are discussed further in \S\ref{sec:interesting}. Within the CSS sample, 13 sources were classified as AGN, 5 as SFGs, and four were unclassified. The location of the CSS sample SFGs are shown on Figures~\ref{fig:rfir} and \ref{fig:qplot} by thick black circles surrounding each source. 

\subsection{Radio-Far Infrared Correlation}
\label{sec:rfirc}
The radio-far infrared correlation (RFIRC) is a well known tight correlation between the far and mid-infrared and radio emission from galaxies (primarily SFGs), first noted in the 1970s, and confirmed in later years \citep[e.g.][]{dickey,dejong}.  \citet{appleton} confirmed the correlation holds to a redshift $z=1$, and it has been recently noted that the RFRIC does not evolve above this redshift, but remains relatively constant to a redshift of $z\sim2$ \citep{frc}. We have utilized the data for both the ELAIS and CDFS field from ATLAS, to further investigate the properties of the sources detected in our MOST observations, and the initial CSS sample (Figure~\ref{fig:rfir}). The solid black line on Figure~\ref{fig:rfir} gives the relationship between radio and infrared flux at 24$\mu$m from \citet{appleton}, where $q_{24}=log(S_{{24\,\mu}{m}}/S_{20\,cm})=0.84$, and the dashed line is the line used by M08 to classify a source as unambiguously AGN (above the dashed line represented by $q_{{24\,\mu}{m}}<-0.16$) or a SFG, AGN or composite object (below the dashed line). Most of our 843\,MHz selected sources are AGN by the M08 classification. The MOST ATLAS sources which fall below this AGN line are further discussed in \S\ref{sec:interesting}. 

The initial CSS sample is also shown in Figure~\ref{fig:rfir}. While these are clearly low-luminosity AGN on the basis of their spectral indices (Figure~\ref{fig:lumalpha}b), they equally clearly fall predominantly in the region occupied by SFGs on the RFIRC. This is an interesting preliminary result which will be investigated further in a follow-up analysis \citep{randall1}. It may be possible that the radio spectral index is steep, betraying the presence of an AGN, but that the AGN contribution to the 1.4\,GHz flux density is small enough (or perhaps is balanced by an equivalent contribution from the AGN to the FIR) that the RFIRC is still obeyed. If this is the case, the location of the CSS sources close to the RFIRC may actually be a consequence of the 1.4\,GHz luminosity being dominated by star formation, and the host galaxy being a composite object. This would be consistent with earlier work suggesting that CSS sources are actively star forming \citep{Review,labiano,morganti,holtb}, and highlight this population as an ideal resource for exploring AGN feedback effects on star formation in galaxies. Only the use of radio discriminants (such as morphology or spectral index) will indicate whether the object hosts an AGN \citep{roy,buriedagn}.

\begin{figure}
\begin{center}
\includegraphics[scale=0.45]{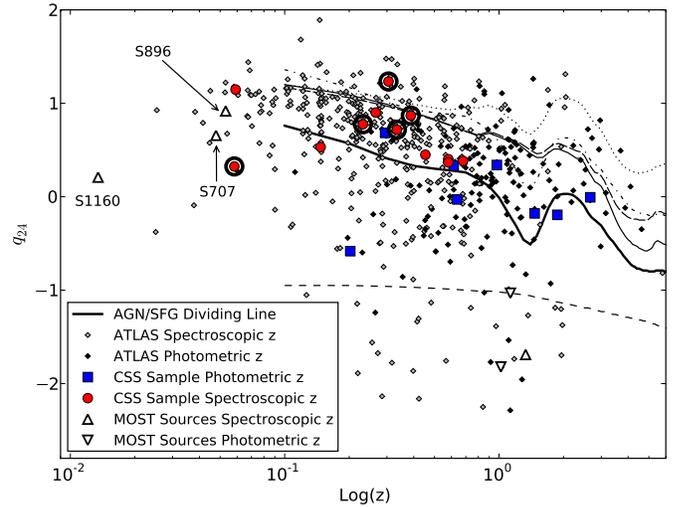}
\caption{$q_{24\mu}$$_{m}$ as a function of redshift for the entire ATLAS catalogue, split into spectroscopic and photometric redshifts. The initial CSS sample is overlaid as the large red circles for those sources with spectroscopic redshifts, and blue squares for those with photometric redshifts. The MOST ATLAS sources with 24$\mu$m detections are shown as the hollow triangles (sources with spectroscopic redshifts) or inverted hollow triangles (sources with photometric redshifts. The SED tracks describe several different models. The heavy dashed line represents an Arp220 type object \citep{dev99}, the thin solid black line is a model for Mrk273 \citep{dev99}, the regular dashed line is a radio-loud QSO \citep{elvis94}, the dotted line is a LIRG model \citep{rieke09}, and the dash-dotted line represents a ULIRG model \citep{rieke09}. The thick black line shows the dividing line between AGN and SFGs from \citet{seymour}. Objects with a thick black circle around the data point are from the CSS sample, and are those spectroscopically classified as SFGs. 
\label{fig:qplot}}
\end{center}
\end{figure}

In Figure~\ref{fig:qplot} the $q_{24}$ values are shown against redshift, with several different spectral energy distribution (SED) tracks from \citet{elvis94,dev99}, and \citet{rieke09} overlaid.  A suggested dividing line between AGN and SFGs from \citet{seymour} is also shown. The CSS sample here are again consistent with having their radio and FIR emission dominated by star formation as are the three nearby MOST ATLAS sources (as labelled in Figures~\ref{fig:rfir} and \ref{fig:qplot}). The SED tracks near these objects include those for Luminous Infrared Galaxies (LIRGs), Ultra-Luminious Infrared Galaxies (ULIRGs), Arp 220 and Mrk 273, two nearby ULIRGs that both host an AGN \citep{engel,iwa11}. The three MOST ATLAS sources in the AGN-dominated region at the bottom of the figure, are likely AGN, and this is supported by the SED of the radio-loud quasar \citep{elvis94} closest to these AGN. Although the SED tracks do not extend to the lowest redshifts of our sample, four of the CSS sample are likely to be LIRGs or ULIRGs, similar to F00183$-$7111, a powerful ULRIG, that has strong radio emission from both an AGN and vigorous star forming activity \citep{foo}, but less luminous in the radio. \citet{foo} have suggested that F00183-7111 is in the earliest phase of the formation of a quasar, where a ``quasar-mode'' AGN \citep{best06,croton} is hosted by a star forming ULIRG, where the star formation is fueled by gas from a past major or minor merger event. Whilst the radio jets of F00183$-$7111 are currently confined by the host galaxy, when they break through this dense gas, the jets will begin the transition of this source into a typical radio-loud quasar, quenching the star formation at the same time \citep{foo}. These CSS sources will be further investigated in a future paper, \citet{randall1}.
 
\subsection{Unusual Sources}
\label{sec:interesting}
We have identified several interesting sources in our final MOST ATLAS catalogue, particularly those sources which lie close to the radio-far infrared correlation (and consequently are likely to host, or even be dominated by, star formation), and one possible ultra-steep spectrum object.

\subsubsection{Possible SFGs in ATLAS}
Of the four single sources that fall close to the RFIRC, three are disk or spiral-like galaxies based on their luminosity profiles in the 3.6$\mu$m SWIRE images (labelled by name in Figures~\ref{fig:rfir}, and \ref{fig:qplot}), which suggests they may be SFGs. The fourth object appears as an elliptical, with no evidence of a disk or interaction, and was not further investigated. Figure~\ref{fig:s707} shows the SWIRE 3.6$\mu$m images with the 1.4\,GHz radio contours overlaid, indicating the clear disk or spiral structure of these three galaxies. Of the three blended sources near the RFIRC, there is less evidence for disks or star-formation in the SWIRE data. For the two closest to the RFIRC (S259/S269 and S1235/S1228/S1230/S1243), the radio emission could possibly still be associated with star formation, but it is harder to constrain without clear evidence of a disk or spiral structure in the host galaxy. We briefly discuss S707, S897 and S1160 below.

\subsubsection{S707 (ATELAIS J003828.02-433847.2)}
\label{sec:s707}
\begin{figure*}
\begin{center}
\includegraphics[scale=0.28]{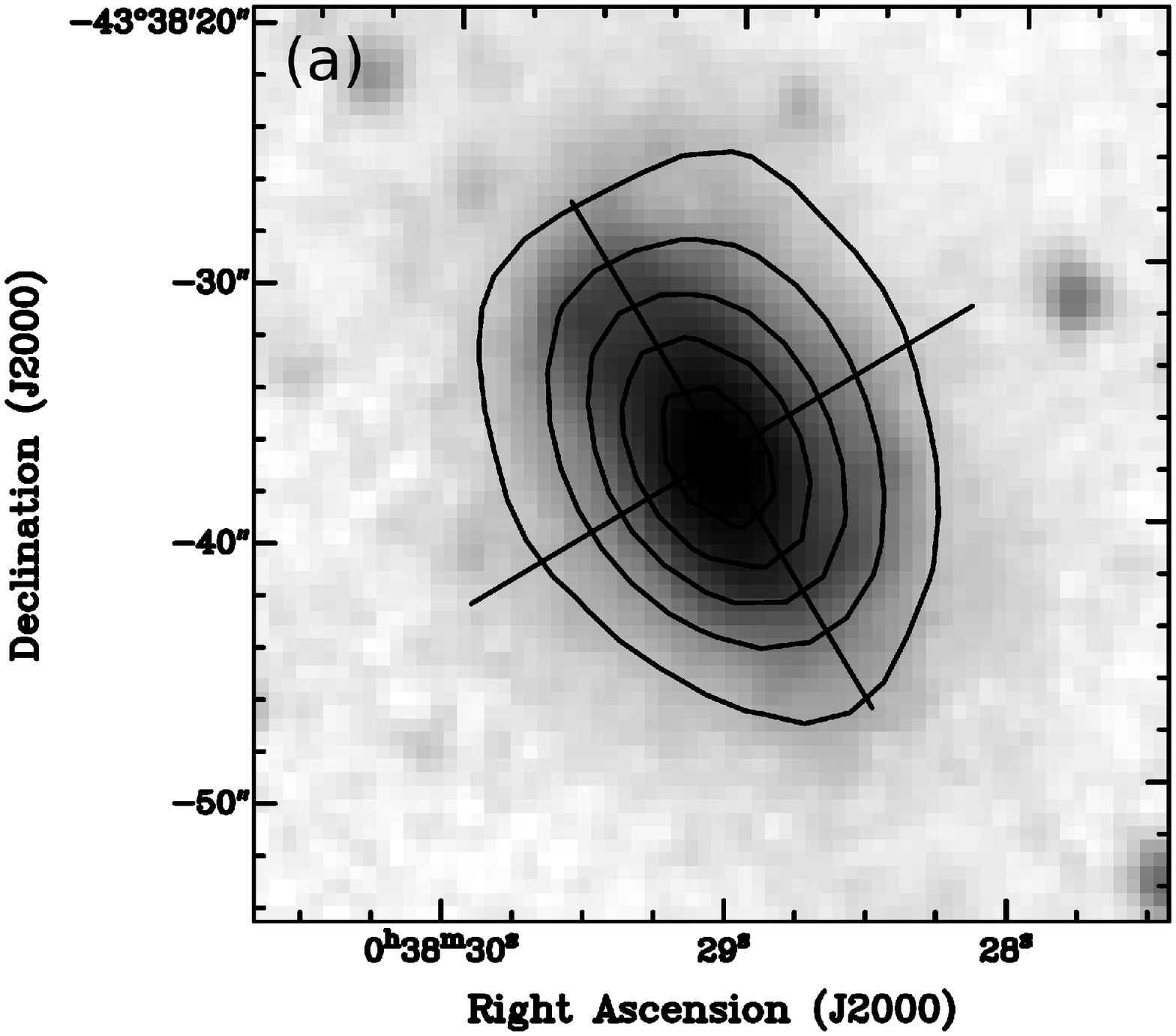}\includegraphics[scale=0.3]{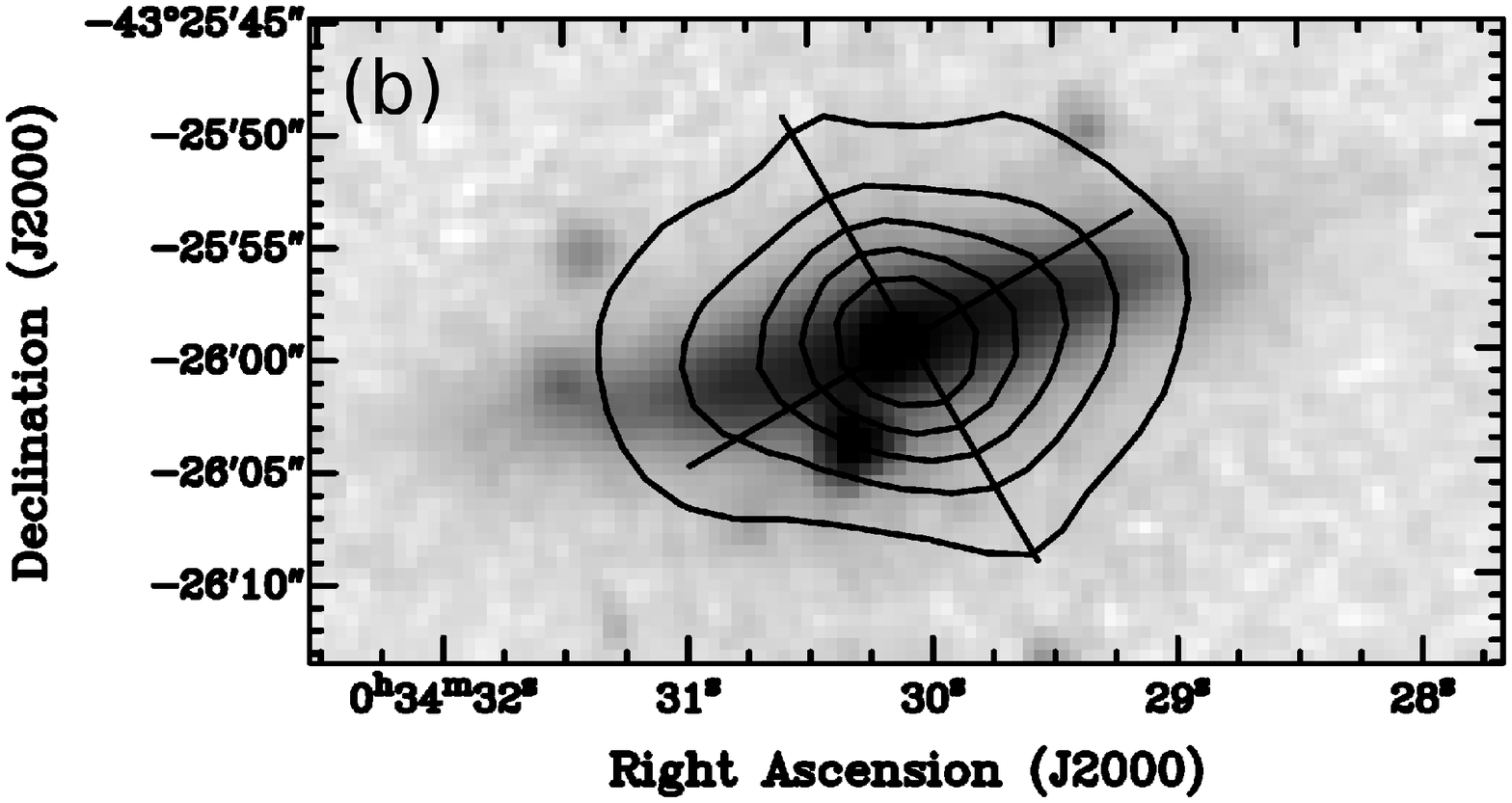}\includegraphics[scale=0.28]{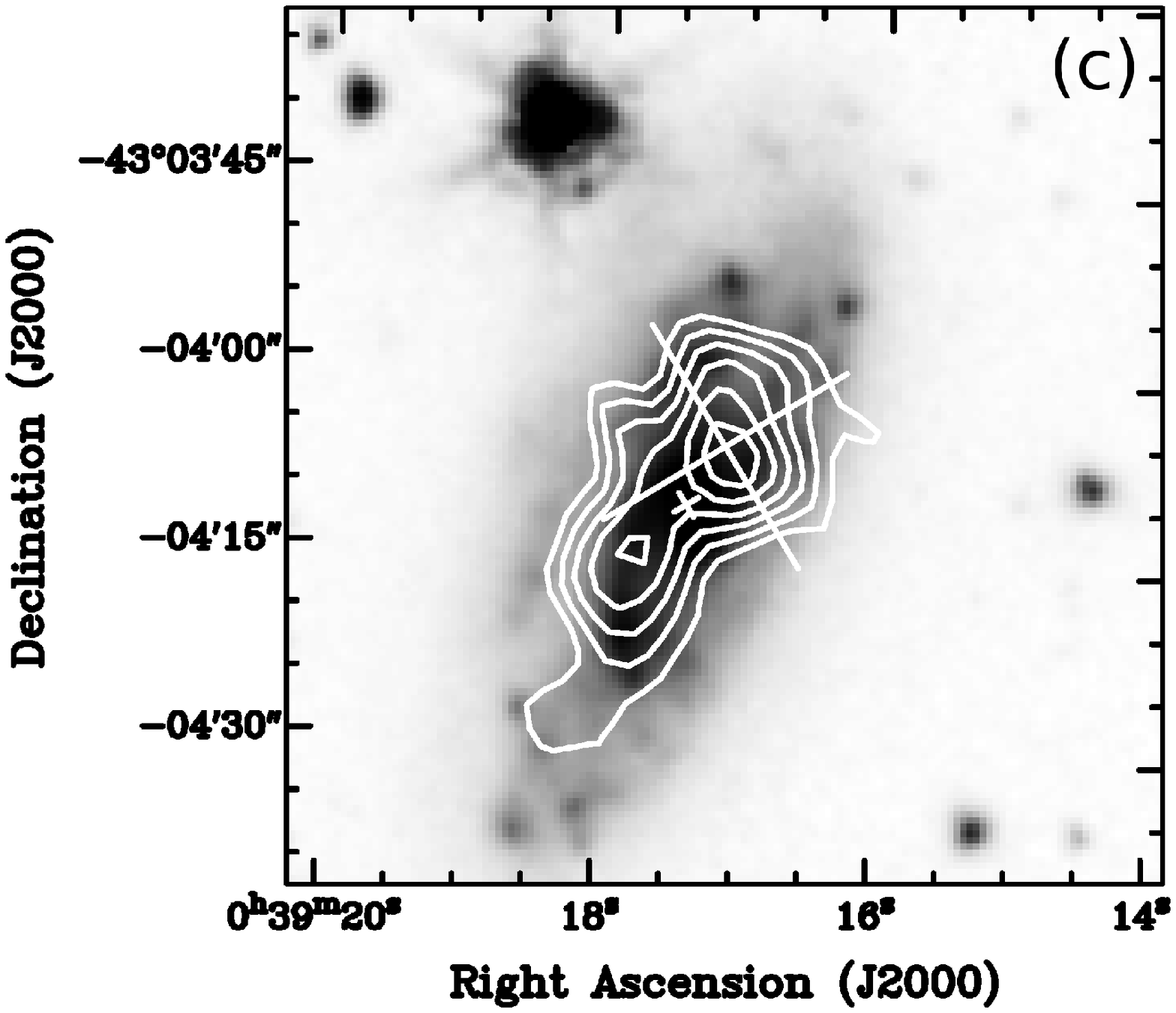}
\caption{(a) 1.4\,GHz radio contours overlaid on the SWIRE 3.6$\mu$m greyscale image for source S707. The radio contours begin at 0.4\,mJy, and increase in increments of 0.4\,mJy to the maximum level of 2\,mJy. The cross is the radio position from M08. (b) 1.4\,GHz radio contours overlaid on the SWIRE 3.6$\mu$m greyscale image for source S896. The radio contours are between 0.1 and 1\,mJy, increasing in increments of 0.2\,mJy. The cross is the radio position from M08. (c) 1.4\,GHz radio contours overlaid on the SWIRE 3.6$\mu$m greyscale image for source S1160. The radio contour levels begin at 0.5\,mJy, and continue up to 1\,mJy, in increments of 0.1\,mJy. The large cross is the radio position from M08, and the small cross in the centre of the galaxy is the SWIRE position.
\label{fig:s707}}
\end{center}
\end{figure*}

S707 is a possible star-forming galaxy at a redshift of $z=0.048$, where the host galaxy has a disk structure, as evidenced by is luminosity profile as seen in Figure~\ref{fig:s707}a. The radio morphology of this object is a single point source at the centre of the SWIRE galaxy, suggesting the low-luminosity AGN in the centre of the galaxy is the primary source of the radio emission, rather than star formation in the disk. Nuclear star formation may also be possible, and the optical spectrum of this object suggests there is ongoing star formation. The radio source has an overall steep spectrum, $\alpha_{fit}=-1.17$, although it is much flatter between 1.4 and 2.3\,GHz ($\alpha^{2.3}_{1.4}=-0.45$), than 843\,MHz and 1.4\,GHz ($\alpha^{1.4}_{0.843}=-1.66$), as expected from an AGN. S707 has a rest-frame luminosity of L$_{1.5}=3.85\times10^{22}$\,WHz$^{-1}$, from the observed convolved 1.4\,GHz flux density, resulting in an upper limit on the star formation rate of $21M_{\sun}$yr$^{-1}$ \citep[assuming all the radio emission is from star formation;][]{sfr}. It also has $q_{{24\,\mu}{m}}=0.57$, suggesting a small fraction of the radio emission is from star formation, whilst the majority is from the AGN. Optically, this source has a similar morphology to the 3.6$\mu$m galaxy, with magnitudes $B=15.8$, $V=15.2$ and $R=14.7$ from M08.

\subsubsection{S896 (ATELAIS J003429.33-432614.4)}
\label{sec:s896}
S896 appears to be hosted by an edge-on disk galaxy, with some possible ongoing star formation that can be seen in Figure~\ref{fig:s707}b, at a redshift of $z=0.053$. Although the radio emission appears to be emitted primarily from the central object, there is most likely some contribution to the overall flux density from star formation in the disk, as the optical spectrum suggests there is some star formation ongoing. This object has a rest-frame luminosity of L$_{1.5}=1.92\times10^{22}$\,WHz$^{-1}$, calculated from the observed convolved 1.4\,GHz flux density. This implies an upper limit on the star formation rate of $\sim11M_{\sun}$yr$^{-1}$, assuming all the radio emission is from star formation. However, it has an overall steep spectral index of $\alpha_{fit}=-1.21$, which suggests that the source of the radio emission is a low-luminosity AGN. For this source $q_{{24\mu}{m}}=0.87$, consistent with the idea that the radio emission is primarily from the AGN. 

\subsubsection{S1160 (ATELAIS J003915.11-430428.5)}
\label{sec:s1160}
S1160 is hosted by a star-forming spiral galaxy, shown in Figure~\ref{fig:s707}c, at a redshift of $z=0.0135$ \citep{jones,jones1}. The radio emission is clearly associated with star formation, as the radio contours trace the star formation in the spiral arms of this galaxy. However, the radio position of the source is offset from the galaxy centre, assigned as the location of the brightest peak in the complex radio structure (M08). The radio luminosity of this source is L$_{1.42}=4.60\times10^{21}$\,WHz$^{-1}$ which results in an upper limit on the star formation rate of $2.5M_{\sun}$yr$^{-1}$ (if all radio emission is due to star formation). The optical morphology of this galaxy is very similar to the SWIRE morphology, with a clear spiral structure. The optical host galaxy has magnitudes from the SuperCOSMOS Science Archive, although these magnitudes should be treated with caution, as it is a large, relatively nearby galaxy and measuring accurate magnitudes in an automated fashion for such sources is difficult due to their complex structure. The SuperCOSMOS magnitudes are $I=13.4$, $R=12.9$ and B$_{J}=7.9$. The optical spectrum is also indicative of star formation. S1160 is detected in all five of the SWIRE bands, with a $q_{24\mu\,m}=0.25$. This $q_{{24\,\mu}{m}}$ value is still consistent with the source being either a SFG or an AGN, and it most likely there is a non-negligible contribution to the radio emission from star formation in this source. The radio flux of this object is interesting, as it has a similar flux density between 843\,MHz and 1.4\,GHz (S$_{0.843}=12.7$\,mJy compared to S$_{1.4}=11.6$\,mJy or S$_{M08}=12.9$\,mJy), but drops rapidly away to S$_{2.3}=5.3$\,mJy. This gives a spectral index $\alpha^{1.4}_{0.843}=-0.2$ and $\alpha^{2.3}_{1.4}=-1.6$, with an overall spectral index of $\alpha_{fit}=-1.03$. This spectral shape is more indicative of AGN activity, but given the radio morphology and the FIR constraints, the radio emission from this object is likely a combination of star formation and AGN activity.  

\subsubsection{S1256 (ATELAIS J003053.26-425215.3): A candidate Ultra-Steep Spectrum source}
\label{sec:s1256}
\begin{figure}
\begin{center}
\includegraphics[scale=0.3,angle=-90]{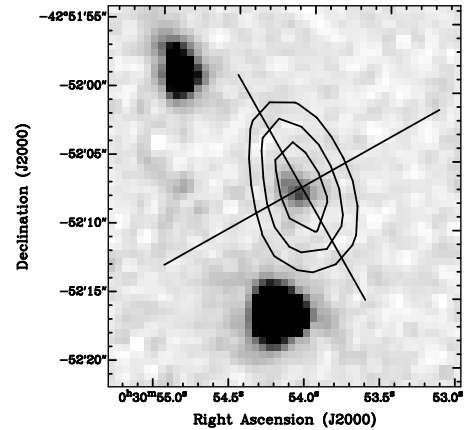}
\caption{1.4\,GHz radio contours overlaid on the SWIRE 3.6$\mu$m greyscale image for source S1256. The radio contours are at levels of 0.5, 0.75 and 1\.mJy. The cross is the radio position from M08.
\label{fig:s1256}}
\end{center}
\end{figure}

High redshift radio galaxies are often found by identifying radio sources with steep or ultra steep radio spectra ($\alpha < -1$), via the well-known $z$-$\alpha$ relation \citep[][and references therein;]{debreuck04,klamer,ishwara}. This trend was first noticed in the early 1980s, when it was found that the fraction of radio sources optically identified on optical plates tended to have the steepest radio spectra. S1256 is an isolated point source, with $S_{0.843}=4.38$\,mJy and $S_{1.4}=0.81$\,mJy, that is not detected in the 2.3\,GHz ATLAS image, with a spectral index $\alpha^{1.4}_{0.843}=-3.33$. Following Z11, using the 3$\sigma$ detection limit of 300$\mu$\,Jy, an upper limit on the spectral index is found to be $\alpha^{2.3}_{1.4}<-2.41$, which suggests that this source is possibly an ultra-steep spectrum source, and therefore at high redshift. Further radio imaging and flux density measurements would be necessary to confirm S1256 as an ultra-steep spectrum source. S1256 has no optical counterpart or confirmed redshift, but is detected in the SWIRE 3.6 and 4.5$\mu$m bands. Its flux densities in these bands are low ($S_{3.6\mu\,m}=17.5\,\mu$Jy and $S_{4.5\mu\,m}=19.9\,\mu$Jy). The infrared morphology is point-like and is shown in Figure~\ref{fig:s1256}. Another possible cause of this steep spectrum is variability; where the radio source has faded away since the 1.4\,GHz observations were completed, and before the 2.3\,GHz observations began. Further observations would also be necessary to confirm whether variability is the cause of the ultra-steep radio spectrum.

\section{Conclusions}
\label{sec:concl}
We have presented a new catalogue of 843\,MHz radio sources, cross-matched to ATLAS at 1.4 and 2.3\,GHz, and explored the properties of this catalogue with spectral index as a function of flux density. Our results do not support the hypothesis that there is a significant flattening of the spectral index with decreasing flux density values. However, we cannot rule out the possibility without further, deep radio data at different radio frequencies. Our analysis of the distribution of steep and flat spectrum sources with redshift, luminosity and infrared flux density indicates that most of the sources in our MOST ATLAS catalogue are AGN, inferred from the spectral index and infrared properties. An initial sample of faint CSS sources in ATLAS has also been selected, with their basic properties explored. We will explore the properties of the CSS sample and the MOST ATLAS selected GPS sources in depth in future work, and compare and contrast this faint sample to our bright sample \citep{randall}. The spectral index properties of ATLAS radio sources across a wide range of frequencies will also be explored in future work, with the aim of distinguishing between the proportions of the populations of low-luminosity, and/or core-dominated AGN and SFGs driving this effect. 

\section*{Acknowledgements}
We would also like to thank our referee, Rogier Windhorst, for his extensive and thorough comments on this paper. We thank the staff at Molonglo for completing the observations and Dick Hunstead for performing the initial data reduction process. We also thank Dick and Elaine Sadler for their input; their comments and help are greatly appreciated. The authors would also like to thank Nick Seymour for the nicely formatted SED templates from various places in the literature for the $q_{24}$ plot.\\
The Australia Telescope Compact Array is part of the Australia Telescope which is funded by the Commonwealth of Australia for operation as a National Facility managed by CSIRO.\\
This research has made use of the NASA/IPAC Extragalactic Database (NED), which is operated by the Jet Propulsion Labratory, Caltech, under contract with NASA.\\

\section*{Supporting Information}
\noindent
An additional table of Supporting Information is available in the online version of this article.
\\

\noindent
\textbf{Table~\ref{table:catalogue}.} The full version of the catalogue, containing all flux density information, ATLAS cross-IDs, optical magnitudes, SWIRE IDs and other relevant information.\\

\noindent
Please note: Wiley-Blackwell are not responsible for the content or functionality of any supporting material supplied by the authors. Any queries (other than missing material) should be directed to the corresponding author for the article.

\bsp

\label{lastpage}


\begin{thebibliography}{99}
\bibitem[\protect\citeauthoryear{Afonso et al.}{2005}]{afonso} Afonso J., Georgakakis A., Almeida C., Hopkins A.~M., Cram L.~E., Mobasher B., Sullivan M., 2005, ApJ, 624, 135
\bibitem[\protect\citeauthoryear{Afonso et al.}{2006}]{afonso1} Afonso J., Mobasher B., Koekemoer A., Norris R.~P., Cram L., 2006, AJ, 131, 1216
\bibitem[\protect\citeauthoryear{Alexander et al.}{2001}]{alexander} Alexander D.~M., et al., 2001, ApJ, 554, 18
\bibitem[\protect\citeauthoryear{Appleton et al.}{2004}]{appleton} Appleton P.~N., et al.,  2004, ApJS, 154, 147
\bibitem[\protect\citeauthoryear{Banfield et al.}{in prep.}]{banfield} Banfield J., et al., in prep.
\bibitem[\protect\citeauthoryear{Best et al.}{2006}]{best06} Best P.~N., Kaiser C.~R., Heckman T.~M., Kauffmann G., 2006, MNRAS, 368L, 67
\bibitem[\protect\citeauthoryear{Berta et al.}{2006}]{berta} Berta S., et al., 2006, A\&A, 451, 881
\bibitem[\protect\citeauthoryear{Berta et al.}{2008}]{berta1} Berta S., et al., 2008, A\&A, 488, 533
\bibitem[\protect\citeauthoryear{Bock et al.}{1999}]{bock} Bock D.~C.-J., Large M.~I., Sadler E.~M., 1999, ApJ, 117, 1578
\bibitem[\protect\citeauthoryear{Burgarella et al.}{2005}]{burgarella} Burgarella D., et al., 2005, AJ, 619, L63
\bibitem[\protect\citeauthoryear{Condon}{1984}]{condon} Condon J.~J., 1984, ApJ, 287, 461
\bibitem[\protect\citeauthoryear{Condon et al.}{1998}]{NVSS} Condon J. J., Cotton W.~D., Greisen E.~W., Yin Q.~F., Perley R.~A., Taylor G.~B., Broderick J.~J., 1998, AJ, 115, 1693
\bibitem[\protect\citeauthoryear{Croton et al.}{2006}]{croton} Croton D.~J., et al., 2006, MNRAS, 365, 11
\bibitem[\protect\citeauthoryear{de Breuck et al.}{2004}]{debreuck04} De Breuck C., Hunstead R.~W., Sadler E.~M., Rocca-Volmerange B., Klamer I., 2004, MNRAS, 347, 837
\bibitem[\protect\citeauthoryear{de Jong et al.}{1985}]{dejong} de Jong T., Klein U., Wielebinski R., Wunderlich E., 1985, A\&A, 147, L6
\bibitem[\protect\citeauthoryear{Devriendt, Guiderdoni \& Sadat}{1999}]{dev99} Devriendt J.~E.~G., Guiderdoni B., Sadat R., 1999, A\&A, 350, 381
\bibitem[\protect\citeauthoryear{Dickey \& Saltpeter}{1984}]{dickey} Dickey J.~M., Saltpeter E.~E., 1984, ApJ, 284, 461
\bibitem[\protect\citeauthoryear{Elvis et al.}{1994}]{elvis94} Elvis M., et al., 1994, ApJS, 95, 1
\bibitem[\protect\citeauthoryear{Engel et al.}{2011}]{engel} Engel H., Davies R.~I., Genzel R., Tacconi L.~J., Sturm E., Downes D., 2011, ApJ, 729, 58
\bibitem[\protect\citeauthoryear{Fanaroff \& Riley}{1974}]{fr} Fanaroff B.~L. \& Riley J.~M., 1974, MNRAS, 167, 31
\bibitem[\protect\citeauthoryear{Fanti}{2009a}]{cfanti} Fanti C., 2009a, Astron. Nachr., 330, 120
\bibitem[\protect\citeauthoryear{Fanti}{2009b}]{rfanti} Fanti R., 2009b, Astron. Nachr., 330, 303
\bibitem[\protect\citeauthoryear{Fanti et al.}{2011}]{fanti} Fanti C., Fanti R., Zanichelli A., Dallacasa D., Stanghellini C., 2011, A\&A, 528, A110
\bibitem[\protect\citeauthoryear{Fazio et al.}{2004}]{fazio} Fazio G.~G., et al., 2004, ApJS, 154, 10
\bibitem[\protect\citeauthoryear{Georgakakis et al.}{1999}]{georg} Georgakakis A., Mobasher B., Cram L., Hopkins A., Lidman C., Rowan-Robinson M., 1999, MNRAS, 306, 708
\bibitem[\protect\citeauthoryear{Gruppioni et al.}{1999}]{grup} Gruppioni C., et al., 1999, MNRAS, 305, 297
\bibitem[\protect\citeauthoryear{Hales et al.}{in prep.}]{chales} Hales C.~A., et al., in prep.
\bibitem[\protect\citeauthoryear{Hill et al.}{1999}]{hill99} Hill T.~L., Heisler C.~A., Sutherland R., Hunstead R.~W., 1999, AJ, 117, 111
\bibitem[\protect\citeauthoryear{Hill et al.}{2001}]{hill01} Hill T.~L., Heisler C.~A., Norris R.~P., Reynolds J.~E., Hunstead R.~W., 2001, AJ, 121, 128
\bibitem[\protect\citeauthoryear{Holt}{2009}]{holtb} Holt J., 2009, Astron. Nachr., 330, 789
\bibitem[\protect\citeauthoryear{Hopkins}{2004}]{sfr} Hopkins A.~M., 2004, AJ, 615, 209
\bibitem[\protect\citeauthoryear{Hopkins et al.}{1998}]{srccnts} Hopkins A.~M., Mobasher B., Cram L., Rowan-Robinson M., 1998, MNRAS, 296, 839
\bibitem[\protect\citeauthoryear{Hopkins et al.}{2002}]{sfind} Hopkins A.~M., Miller C.~J., Connolly A.~J., Genovese C., Nichol R.~C., Wasserman L., 2002, AJ, 123, 1086
\bibitem[\protect\citeauthoryear{Hopkins et al.}{2003}]{phoenix} Hopkins A.~M., Afonso J., Chan B., Cram L.~E., Georgakakis, A., Mobasher B., 2003, AJ, 125, 465
\bibitem[\protect\citeauthoryear{Ibar et al.}{2009}]{ibar} Ibar E., Ivison R.~J., Biggs A.~D., Lal D.~V., Best P.~N., Green D.~A., 2009, MNRAS, 397, 281
\bibitem[\protect\citeauthoryear{Ibar et al.}{2010}]{ibar1} Ibar E., Ivison R.~J., Best P.~N., Coppin K., Pope A., Smail I., Dunlop J.~S., 2010, MNRAS, 401L, 53
\bibitem[\protect\citeauthoryear{Ishwara-Chandra et al.}{2010}]{ishwara} Ishwara-Chandra C.~H., Sirothia S.~K., Wadadekar Y., Pal S., Windhorst R, 2010, MNRAS, 405, 436
\bibitem[\protect\citeauthoryear{Iwasawa et al.}{2011}]{iwa11} Iwasawa K., et al., 2011, A\&A, 528A, 137
\bibitem[\protect\citeauthoryear{Jones et al.}{2004}]{jones} Jones D.~H., et al., 2004, MNRAS, 355, 747
\bibitem[\protect\citeauthoryear{Jones et al.}{2009}]{jones1} Jones D.~H., et al., 2009, MNRAS, 399, 683
\bibitem[\protect\citeauthoryear{Kellermann}{1964}]{kellermann} Kellermann K., 1964, ApJ, 140, 969
\bibitem[\protect\citeauthoryear{Kesteven, Bridle \& Brandie}{1977}]{kesteven} Kesteven M.~J.~L., Bridle A.~H., Brandie G.~W., 1977, AJ, 82, 541
\bibitem[\protect\citeauthoryear{Klamer et al.}{2006}]{klamer} Klamer I.~J., Ekers R.~D., Bryant J.~J., Hunstead R.~W., Sadler E.~M., De Breuck C., 2006, MNRAS, 371, 852
\bibitem[\protect\citeauthoryear{Komatsu et al.}{2011}]{wmap} Komatsu E., et al., 2011, ApJS, 192, 18
\bibitem[\protect\citeauthoryear{Kunert-Bajraszewska \& Marecki}{2007}]{kunert07} Kunert-Bajraszewska M., Marecki A., 2007, A\&A, 469, 437
\bibitem[\protect\citeauthoryear{Labiano et al.}{2008}]{labiano} Labiano A., O'Dea C.~P., Barthel P.~D., de Vries W.~H., Baum S.~A., 2008, A\&A, 477, 491
\bibitem[\protect\citeauthoryear{Large et al.}{1981}]{large} Large M.~I., Mills B.~Y., Little A.~G., Crawford D.~F., Sutton J.~M, 1981, MNRAS, 194, 693
\bibitem[\protect\citeauthoryear{Longair}{1966}]{longair} Longair M.~S., 1966, MNRAS, 133, 421
\bibitem[\protect\citeauthoryear{Lonsdale et al.}{2003}]{lonsdale} Lonsdale C.~J., et al., 2003, PASP, 115, 897
\bibitem[\protect\citeauthoryear{Lonsdale et al.}{2004}]{lonsdale1} Lonsdale, C.~J., et al., 2004, ApJS, 154, 54
\bibitem[\protect\citeauthoryear{Magliocchetti et al.}{2000}]{maglio} Magliocchetti M., Maddox S.~J., Wall J.~V., Benn C.~R., Cotter G, 2000, MNRAS, 318, 1047
\bibitem[\protect\citeauthoryear{Mao et al.}{2010}]{minnie} Mao M.~Y., Sharp R., Saikia D.~J., Norris R.~P., Johnston-Hollitt M., Middelberg E., Lovell J.~E.~J.., 2010, MNRAS, 406, 2578
\bibitem[\protect\citeauthoryear{Mao et al.}{2011}]{frc} Mao M.~Y., Huynh M.~T., Norris R.~P., Dickinson M., Frayer D., Helou G., Monkiewicz J.~A., 2011, ApJ, 731, 79
\bibitem[\protect\citeauthoryear{Mao et al.}{in prep.}]{mao} Mao M.~Y., et al., in prep.
\bibitem[\protect\citeauthoryear{Martin et al.}{2005}]{martin} Martin D.~C., et al., 2005, AJ, 619, L1
\bibitem[\protect\citeauthoryear{Mauch et al.}{2003}]{SUMSS2} Mauch T., Murphy T., Buttery H.~J., Curran J., Hunstead R.~W., Piestrzynski B., Robertson J.~G., Sadler E.~M., 2003, MNRAS, 342, 1117
\bibitem[\protect\citeauthoryear{Middelberg et al.}{2008}]{Middelberg08} Middelberg E., et al. (M08), 2008, AJ, 135, 1276 
\bibitem[\protect\citeauthoryear{Middelberg et al.}{2011}]{ifrs} Middelberg E., Norris R.~P., Hales C.~A., Seymour N., Johnston-Hollitt M., Huynh M.~T., Lenc E., Mao M.~Y., 2011, A\&A, 526A, 8
\bibitem[\protect\citeauthoryear{Mills}{1981}]{mills} Mills B.~Y., 1981, PASA, 4, 156
\bibitem[\protect\citeauthoryear{Morganti et al.}{2009}]{morganti} Morganti R., Emonts B., Holt J., Tadhunter C., Oosterloo T., Struve C., 2009, Astron. Nachr., 330, 789
\bibitem[\protect\citeauthoryear{Murphy et al.}{2010}]{at20gcat} Murphy T. et al., 2010, MNRAS, 402, 2403
\bibitem[\protect\citeauthoryear{Norris et al.}{2006}]{ray} Norris R.~P., et al. (N06), 2006, AJ, 132, 2409
\bibitem[\protect\citeauthoryear{Norris, Middelberg \& Boyle}{2007}]{buriedagn} Norris R.~P., Middelberg E., Boyle B.~J., 2007, ASPC, 380, 229
\bibitem[\protect\citeauthoryear{Norris et al.}{2011}]{foo} Norris R.~P., Lenc E., Roy A.~L., Spoon H., 2011, preprint (arXiv: 1107.3895N)
\bibitem[\protect\citeauthoryear{O'Dea}{1998}]{Review} O'Dea C., 1998, PASP, 110, 493
\bibitem[\protect\citeauthoryear{Oliver et al.}{2000}]{oliver} Oliver S.,  Rowan-Robinson M., Alexander D.~M., et al., 2000, MNRAS, 316, 749
\bibitem[\protect\citeauthoryear{Oort \& Windhorst}{1985}]{oort} Oort M.~J.~A., Windhorst R.~A., 1985, A\&A, 145, 405
\bibitem[\protect\citeauthoryear{Owen \& Morrison}{2008}]{owen} Owen F.~N., Morrison G.~E., 2008, AJ, 136, 1889
\bibitem[\protect\citeauthoryear{Owen et al.}{2009}]{owen1} Owen F.~N., Morrison, G.~E., Klimek M.~D., Greisen E.~W., 2009, AJ, 137, 4846
\bibitem[\protect\citeauthoryear{Polatidis \& Conway}{2003}]{young} Polatidis, A., \& Conway, A., 2003, PASA, 20, 69
\bibitem[\protect\citeauthoryear{Prandoni et al.}{2000a}]{prandoni} Prandoni I., Gregorini L., Parma P., de Ruiter H.~R., Vettolani G., Wieringa M.~H., Ekers R.~D., 2000a, A\&AS, 146, 31
\bibitem[\protect\citeauthoryear{Prandoni et al.}{2000b}]{prandoni1} Prandoni I., Gregorini L., Parma P., de Ruiter H.~R., Vettolani G., Wieringa M.~H., Ekers R.~D., 2000b, A\&AS, 146, 41
\bibitem[\protect\citeauthoryear{Prandoni et al.}{2006}]{prandoni2} Prandoni I., Parma P., Wieringa M.~H., de Ruiter H.~R., Gregorini L., Mignano A., Vettolani G., Ekers R.~D., 2006, A\&A, 457, 517
\bibitem[\protect\citeauthoryear{Prandoni et al.}{2011}]{prandoni4} Prandoni I., Bernardi G., Di Vincenzo A., de Bruyn A.~G., 2011, submitted to A\&A
\bibitem[\protect\citeauthoryear{Puccetti et al.}{2006}]{puccetti} Puccetti S., et al., 2006, A\&A, 457, 501
\bibitem[\protect\citeauthoryear{Randall et al.}{2011}]{randall} Randall K.~E., Hopkins A.~M., Norris R.~P, Edwards P.~G., 2011a, MNRAS, 416, 1135
\bibitem[\protect\citeauthoryear{Randall et al.}{in prep.}]{randall1} Randall K.~E., et al., in prep.
\bibitem[\protect\citeauthoryear{Rieke et al.}{2004}]{rieke} Rieke G.~H., Young E.~T., Engelbracht C.~W., et al., 2004, ApJS, 154, 25
\bibitem[\protect\citeauthoryear{Rieke et al.}{2009}]{rieke09} Rieke G.~H., Alonso-Herrero A., Weiner B.~J., P{\'e}rez-Gonz{\'a}lez P.~G., Blaylock M., Donley J.~L., Marcillac D., 2009, ApJ, 692, 556
\bibitem[\protect\citeauthoryear{Robertson}{1991}]{robertson} Robertson J.~G., 1991, Aust. J. Phys., 44, 729
\bibitem[\protect\citeauthoryear{Rowan-Robinson et al.}{2004}]{rowan} Rowan-Robinson M., Lari C., Perez-Fournon I., et al., 2004, MNRAS, 351, 1290
\bibitem[\protect\citeauthoryear{Roy et al.}{1998}]{roy} Roy A.~L., Norris R.~P., Kesteven M.~J., Troup E.~R., Reynolds J.~E., 1998, MNRAS, 301, 1019
\bibitem[\protect\citeauthoryear{Schlegel, Finkbeiner \& Davis}{1998}]{schlegel} Schlegel D.~J., Finkbeiner D.~P., Davis M., 1998, ApJ, 500, 525
\bibitem[\protect\citeauthoryear{Scoville et al.}{2007}]{scoville} Scoville N., et al., 2007, ApJS, 172, 1
\bibitem[\protect\citeauthoryear{Seymour et al.}{2008}]{seymour} Seymour N., et al., 2008, MNRAS, 386, 1695
\bibitem[\protect\citeauthoryear{Smol{\v c}i{\'c} et al.}{2008}]{smolcic} Smol{\v c}i{\'c} V., et al., 2008, ApJS, 177, 14
\bibitem[\protect\citeauthoryear{Snellen et al.}{1998}]{faintgps} Snellen I.~A.~G., Schilizzi R.~T., de Bruyn A.~G., Miley G.~K., Rengelink R.~B., R{\"o}ttgering H.~J., Bremer M.~N., 1998, A\&AS, 131, 435
\bibitem[\protect\citeauthoryear{Snellen et al.}{1999}]{evolution} Snellen I.~A.~G., Schilizzi R.~T., Miley G.~K., Bremer M.~N., R{\"o}ttgering H.~J.~A., van Langevelde H.~J., 1999, NewAR, 43, 675
\bibitem[\protect\citeauthoryear{Snellen et al.}{2000}]{snell00} Snellen I.~A.~G., Schilizzi R.~T., van Langevelde H.~J, 2000, MNRAS, 319, 429
\bibitem[\protect\citeauthoryear{Tschager et al.}{2003a}]{faint} Tschager, W., Schilizzi R.~T., R{\"o}ttgering H.~J.~A., Snellen I.~A.~G., Miley G.~K., Perley R.~A., 2003a, PASA, 20, 75
\bibitem[\protect\citeauthoryear{Tschager et al.}{2003b}]{tschager} Tschager, W., Schilizzi R.~T., R{\"o}ttgering H.~J.~A., Snellen I.~A.~G., Miley G.~K., Perley R.~A., 2003b, A\&A, 402, 171
\bibitem[\protect\citeauthoryear{Vaccari et al.}{2005}]{vaccari} Vaccari M., et al., 2005, MNRAS, 358, 397
\bibitem[\protect\citeauthoryear{Wilson et al.}{2011}]{cabb} Wilson W.~E., et al., 2011, MNRAS, 416, 832
\bibitem[\protect\citeauthoryear{Windhorst et al.}{1985}]{windhorst85} Windhorst R.~A., Miley G.~K., Owen F.~N., Kron R.~G., Koo D.~C., 1985, ApJ, 289, 494
\bibitem[\protect\citeauthoryear{Windhorst et al.}{1993}]{windhorst} Windhorst R.~A., Fomalont E.~B., Partridge R.~B., Lowenthal J.~D., 1993, AJ, 405, 498
\bibitem[\protect\citeauthoryear{Zinn et al.}{in prep.}]{zinn} Zinn P-C., et al., in prep.
\end{thebibliography}
\end{document}